\documentclass[traditabstract]{aa}                                   \usepackage{graphicx}
\usepackage{txfonts}
\usepackage{natbib}
\usepackage{subfig} \usepackage{enumerate}  \usepackage{amssymb}
\usepackage{lscape}
\usepackage{rotating}
\usepackage{balance}

\newcommand{\mi}{$M_{F814W}$~}
\newcommand{\mb}{$M_{F435W}$~}
\newcommand{\mbi}{$M_{F435W}$-$M_{F814W}$~}

\newcommand{\reff}{$r_{\rm{eff}}$~}

\newcommand{\lir}{$L_{\rm{IR}}$~}
\newcommand{\ha}{H$\alpha$~}
\newcommand{\hb}{H$\beta$~}
\newcommand{\lha}{$L_{\rm{H}\alpha}$}
\newcommand{\ergs}{erg s$^{-1}$~}
\newcommand{\msun}{M$_{\odot}$~}
\newcommand{\mdyn}{M$_{\rm{dyn}}$~}
\newcommand{\lsun}{$L_{\odot}$~}
\newcommand{\hii}{H{\scriptsize{II}}~}
\newcommand{\hi}{H{\scriptsize{I}}~}
\newcommand{\nii}{[N{\scriptsize{II}}]~}
\newcommand{\oiii}{[O{\scriptsize{III}}]~}
\newcommand{\ciii}{[C{\scriptsize{III}}]~}
\newcommand{\civ}{[C{\scriptsize{IV}}]~}
\newcommand{\ld}{$D_{\rm{L}}$~}
\newcommand{\av}{$A_V$~}
\newcommand{\twospace}{$\!\!$}
\newcommand{\onespace}{$\!$}
\newcommand{\tdgs}{TDGs~}
\newcommand{\tdg}{TDG~}
\newcommand{\hst}{\textit{HST}~}

\makeatletter
\newcounter{subsubsubsection}[subsubsection]
\renewcommand\thesubsubsubsection{\thesubsubsection .\@arabic\c@subsubsubsection}
\newcommand\subsubsubsection{\@startsection{subsubsubsection}{4}{\z@}                                     {-3.25ex\@plus -1ex \@minus -.2ex}                                     {1.5ex \@plus .2ex}                                     {\normalfont\normalsize}}
\newcommand*\l@subsubsubsection{\@dottedtocline{3}{10.0em}{4.1em}}
\newcommand*{\subsubsubsectionmark}[1]{}
\makeatother

\begin{document}
   \title{Extranuclear \ha\twospace-emitting complexes in low-z (U)LIRGs: Precursors of tidal dwarf galaxies?}

   \author{D. Miralles-Caballero,
          \inst{1}           L. Colina,
          \inst{1}
	\and
          S. Arribas
	\inst{1}
                    }

       \institute{Departamento de Astrof{\'{\i}}sica, Centro de Astrobiolog{\'{\i}}a, CSIC-INTA, Ctra. de Torrej\'on Ajalvir km 4,
	Torrej\'on de Ardoz, 28850 Madrid, Spain\\
		\email{mirallescds@cab.inta-csic.es}
             }

   \date{Received 7 June 2011; accepted 6 October 2011  }

  \abstract
                    {
   
This paper characterizes the physical and kinematic properties of external massive star-forming regions in a sample of (U)LIRGs. We use high angular resolution ACS images from the \textit{Hubble Space Telescope} (\hst\twospace) in \textit{F435W} ($\sim$ \textit{B}) and \textit{F814W} ($\sim$ \textit{I}) bands, as well as  \ha\twospace-line emission maps obtained with integral field spectroscopy. We find 31 external \ha\onespace-emitting (i.e., young star-forming) complexes in 11 (U)LIRGs. These complexes have in general similar sizes (from few hundreds of pc to about 2 kpc),  luminosities (\mb$<$ -10.65 and L(\ha\twospace)$>$10$^{39}$\lsun\twospace), and metallicities (\mbox{12 + log (O/H) $\sim$ 8.5-8.7}) to extragalactic giant \hii regions and \tdg candidates found in less luminous mergers and compact groups of galaxies. We assess the mass content and the likelihood of survival as \tdgs of the 22 complexes with simple structures in the \hst images based on their photometric, structural, and kinematic properties.  The dynamical tracers used (radius-$\sigma$ and luminosity-$\sigma$ diagrams) indicate that most of the complexes might be self-gravitating entities. The resistance to forces from the parent galaxy is studied by considering the tidal mass of the candidate and its relative velocity with respect to the parent galaxy. After combining the results of previous studies of \tdg searches in ULIRGs a total of 9 complexes satisfy most of the applied criteria and thus show a \textit{high-medium} or \textit{high} likelihood of survival, their total mass likely being compatible with that of dwarf galaxies. They are defined as \tdg candidates. We propose that they probably formed more often during the early phases of the interaction. Combining all data for complexes with IFS data where a significant fraction of the system is covered, we infer a TDG production rate of 0.3 candidates with the highest probabilities of survival per system for the (U)LIRGs class. This rate, though, might decrease to 0.1 after the systems in (U)LIRGs have evolved for 10 Gyr, for long-lived \tdgs\twospace, which would imply that no more than 5-10\% of the overall dwarf population could be of tidal origin.}

\authorrunning{D. Miralles-Caballero et al.}
\titlerunning{Extranuclear \ha complexes in low-z (U)LIRGs: Precusors of TDGs?}
   \maketitle

\section{Introduction}

The question of how galaxies form has been a key issue in extragalactic astronomy for the past few decades. One product of galaxy formation observable primarily in the local universe is the dwarf galaxy population, that usually forms in tidal tails during interactions of giant galaxies up to the present day. These galaxies are now generally called tidal dwarf galaxies (TDGs;~\citealt{Duc98}). 

The scenario of low-mass galaxy formation during giant galaxy collisions was first proposed by~\cite{Zwicky56} despite a lack of strong observational evidence, which was provided later (\citealt{Schweizer78}). Subsequent observations (\citealt{Mirabel91,Hernquist92a}) and numerical simulations (\citealt{Barnes92b,Elmegreen93}) demonstrated that the existence of dwarf galaxies as a result of a major interaction is possible. Since then and up until now, several observing campaigns have been launched in interacting systems, such as those in NGC 2782 (\citealt{Yoshida94}), Arp 105 (\citealt{Duc94,Duc97}), NGC 7252 (\citealt{Hibbard94}), NGC 5291 (\citealt{Duc98,Higdon06}), Arp 245 (\citealt{Duc00}), small samples (\citealt{Weilbacher00,Weilbacher03,Knierman03,Monreal07}), and Arp 305 (\citealt{Hancock09}). \cite{Hibbard95} also developed a successful dynamical N-body model of NGC 7252. Some other investigations have tried to find \tdgs specifically in compact galaxy groups (e.g.,~\citealt{Hunsberger96,Iglesias-Paramo01,Temporin03,Nishiura02}).

All spectroscopic observations of \tdgs have shown that the luminosity-metallicity correlation found for normal dwarf galaxies does not hold for \tdgs\twospace. While dwarf galaxies have lower oxygen abundances at lower luminosities, \tdgs have an approximately constant metallicity of around one-third of the solar value as determined from gaseous emission lines (see e.g.,~\citealt{Duc00,Weilbacher03}). 

The identification of the observed \tdg candidates as real galaxies is not straightforward. They are
embedded in a tidal tail and, most probably, do not have massive dark matter halos. Tidal forces produced by the parent galaxy disturb their gravitational field, strong star formation might blow away the recently accreted gas, and some of the \tdgs may even fall back into the central merger (\citealt{Hibbard95}). An accepted definition that tries to ensure that only the objects that we refer to as tidal dwarf galaxies deserve the classification of a ``galaxy'' is: A tidal dwarf galaxy is a self-gravitating entity of dwarf-galaxy mass built from the tidal material expelled during interactions (\citealt{Duc00,Weilbacher01}).

Taking into account evaporation and fragmentation processes as well as tidal disruption, \cite{Duc04} suggested that a total mass of as high as 10$^9$~\msun may be necessary for it to become a long-lived \tdg\twospace. This is the typical mass of the giant HI accumulations observed near the tip of several long tidal tails. Less massive condensations may evolve, if they survive, into objects more similar to globular clusters. However, this mass criterion is not well-established, and objects with a total mass of a few 10$^7$-a few 10$^9$~\msun are normally considered as \tdg candidates.

Interacting systems (and remnants) constitute the optimal environment for finding TDGs. Although many studies devoted to searching and characterizing TDGs in nearby interacting systems have been carried out in the past few decades, only a few searches for \tdg candidates have been performed to date in the most energetic interacting systems in the local universe, luminous (LIRGs) and  ultraluminous (ULIRGs) infrared galaxies.

The LIRGs and ULIRGs are objects with infrared luminosities of $10^{11} L_{\odot} \leq L_{bol} \sim L_{IR}[8 - 1000 \mu m]$ $< 10^{12} L_{\odot}$ and  $10^{12} L_{\odot} \leq L_{IR}[8 - 1000 \mu m]$\footnote{For simplicity, we define the infrared luminosity to be \mbox{\lir(\lsun) $\equiv$ log ($L_{IR}[8 - 1000 \mu m]$) }.} $< 10^{13} L_{\odot}$, respectively \citep{Sanders96}. The main source of this luminosity in most of them is thought to be  starburst activity~\citep{Genzel98a,Farrah03,Yuan10}. These galaxies are rich in gas and dust, and more than 50\% of LIRGs and most ULIRGs show signs of being involved in a major interaction/merger  \citep[e.g][]{Surace98,Surace00,Cui01,Farrah01,Bushouse02,Evans02,Veilleux06}. 

On the basis of previous studies of nearby interacting galaxies, TDGs are normally identified as extreme star-forming regions in tidal tails. Owing to the high star-formation rates measured in (U)LIRGs and the interacting nature of many of them, we expect to find a significant number of candidates in this environment. Previous studies of TDGs in ULIRGs include: one of the Superantennae (\citealt{Mirabel91}), whose TDG candidate turned out to be a background object (\citealt{Mirabel91}); that of four ULIRGs, where only three TDG candidates were identified (\citealt{Mihos98}); and another of nine ULIRGs where, in five of them, a total of twelve TDG candidates were identified based on their \ha\twospace-emission (\citealt{Monreal07}). This third work also studied the likelihood that the \tdg candidates will survive, and found that five out of the twelve show high probabilities of their remaining a TDG. This paper extends the previous study of \cite{Monreal07} by including galaxies in the LIRG class.

A study of TDGs in the local universe is issential to aid our understanding of galaxy formation at high-z. If TDGs, which form from tidal debris, are long-lived, they could contribute significantly to the total population of dwarf satellites, in addition to primordial dwarfs. Their statistical properties would then have to be modified and the available constraints on cosmological models would have to be updated. Early studies showed that this is the case, even that the overall dwarf population could be of tidal origin (\citealt{Okazaki00}). However, it has been claimed that only a marginal fraction (less than 10\%) of dwarf galaxies in the local universe could have a tidal origin (\citealt{Bournaud06,Wen11,Kaviraj11}). The (U)LIRGs are major contributors to the star-formation rate density at z$\sim$1-2~\citep{Perez-Gonzalez05}. In addition, the observed properties of (U)LIRGs share many
similarities with populations of star-forming galaxies at higher redshifts~\citep{Chapman03,Frayer03,Frayer04,Engel10}. Hence, the study of the local (U)LIRG population provides an opportunity to link the properties of high-z galaxies with those we observe in the nearby universe. In particular, they can be the key to understanding how dwarf galaxies form in the early Universe.

In this paper, we characterize the extranuclear star-forming  regions of a sample of local (z $<$ 0.1) (U)LIRGs using information derived from integral field spectroscopy (IFS) data together with that of high resolution images from the \textit{Hubble Space Telescope} (\hst\twospace) in the \textit{B} and \textit{I} photometric band. We perform a dedicated search for potential \tdg candidates in local (U)LIRGs. We derive and compare candidate properties such as the metallicity, age, and mass of the young stellar population, \ha luminosity, dynamical mass, velocity dispersion, relative velocity etc., with those of \tdg candidates in the literature.  We use these properties to estimate the likelihood of the survival of these regions as future \tdgs\twospace.

The paper is organized as follows. We first describe the sample in section~\ref{sec:sample}. The photometric and spectroscopic data are presented in section~\ref{sec:data}. We then identify and characterize the luminosities, colors, metallicities, etc. of extranuclear \ha\twospace-emitting complexes in~\ref{sec:results}. In section~\ref{sec:discussion}, we select the complexes that are most likely to constitute \tdg candidates and discuss their total mass content and the likeliness of the survival of these regions as future \tdgs\twospace, and the implications for the formation of \tdgs at high-z. Finally, we draw our conclusions in section~\ref{sec:conclusions}.

\section{The sample}
\label{sec:sample}

The galaxies selected for this study were taken from a representative sample of 32 low-z both luminous and ultraluminous infrared galaxies, (U)LIRGs, presented in~\cite{Miralles-Caballero11} (hereafter MC11). It covers the luminosity range 11.39 $\leq$ \lir $\leq $ 12.54 and the distance range from  65 to about 550 Mpc. It also spans all types of nuclear activity, with different excitation mechanisms such as LINER (i.e., shocks, strong winds, weak AGN), HII (star formation), and Seyfert-like activity (presence of an AGN). Finally, all the different morphologies usually identified in these systems are also sampled. 

The main purpose of the paper is to combine photometric and spectroscopic data to perform a physical  characterization of the most luminous extranuclear \ha\twospace-emitting regions in these galaxies. Therefore, we selected systems for which IFS data was available at the moment of analysis, that is a total of 27 systems. As explained in section~\ref{sub:identification}, regions of interest were selected as any high surface brightness compact region (\ha clumps) in the emission line maps (obtained from the IFS data) at a  projected distance from the nucleus of the galaxy greater than 2.5 kpc, and associated with one or several of the star-forming regions (knots) in the images taken with the Advanced Camera for Surveys (ACS), onboard the Hubble Space Telescope (\hst\twospace). On the basis of these criteria \ha\twospace-emitting complexes have been identified in 11 systems. We did not identify other complexes in the remaining sample basically for two reasons: (1) based on the above criteria, in some systems knots are detected but no \ha emission is found in the outskirts in the \ha line map (e.g., in IRAS 22491-1808); (2) the field of view (FoV) covered by the IFS data is not large enough to encompass the full extent of some systems. Hence, the nature of  blue knots along the tidal tail and at its tip, where TDGs are normally found, cannot be assessed using the IFS data. 

Therefore, the final sample consists of 11 systems with at least one region of interest (see table~\ref{table:phot_prop}), or more precisely 7 LIRGs and 4 ULIRGs. For completeness, we also include in our tables the results of~\cite{Monreal07} for IRAS 16007+3743, for which an image in the \textit{B} photometric band was unavailable. According to the classification scheme defined in MC11, IRAS 04315-0840, IRAS 08355-4944, IRAS F10038-3338, and IRAS 15250+3609 are in an advanced phase of a merger (merger phase) where the nuclei have apparently coalesced. However, the tails are still recognizable in the images, in some cases very prominently. The rest of the systems are in earlier stages of the merging process, first contact of pre-merger phases, according to the classification scheme we use.

\section{The data}
\label{sec:data}

The results presented in this paper are based on photometric and spectroscopic data. In the following, we give the details of both sets.

\subsection{\hst imaging}

We retrieved high angular resolution archival \hst images taken with ACS in two broad-band filters, \textit{F814W} and \textit{F435W}. The former is equivalent to the ground-based Johnson-Cousins \textit{I} and the latter differs from the ground-based Johnson-Cousins \textit{B} between 7\% and 12\% in flux~\citep{Sirianni05}. The pixel size for this instrument is 0.05\arcsec\onespace, and given the FoV of the instrument (200\arcsec $\times$ 200\arcsec) the galaxies and their tails in the \textit{I} photometric band are completely covered. More details about the observations and the post-calibration applied can be found in MC11.

Complementary images from the Near Infrared Camera and Multi Object Spectrometer (NICMOS), onboard \hst\twospace, were also retrieved. They were taken with the \textit{F160W} (\textit{H}) band filter. For IRAS 06076-2139, we used an available H band image taken with the Wide Field Camera 3 (WFC3). These \textit{H} images helped us determine effective radii of the galaxies.

\subsection{Integral field spectroscopy maps}

We obtained IFS observations for the galaxies of the sample with two different fiber-based optical integral field systems:

\begin{itemize}
 \item The ULIRGs in our sample were observed with INTEGRAL~\citep{Arribas98}, connected to WYFFOS~\citep{Bingham94} and mounted on the ~\mbox{4.2 m} William Herschel Telescope. The FoV varies depending on the INTEGRAL dithering and pointings used, though in general it is around \mbox{16\arcsec $\times$ 12.3\arcsec}. The angular sampling is 0.90\arcsec per fiber. Details of the data reduction and calibration can be found in~\cite{Garcia-Marin09a} and references therein. \\
 \item The LIRGs in our sample were observed with VIMOS~\citep{leFevre03}, mounted on the Nasmyth focus B on the VLT. The FoV covers \mbox{27\arcsec $\times$ 27\arcsec} with an angular sampling of 0.67\arcsec per fiber. Details of the data reduction and calibration can be found in both~\cite{Arribas08} and~\cite{Rodriguez-Zaurin10}. 
\end{itemize}
 
In both sets of data, the H$\alpha$ line lies within the spectral range and, once calibrated, H$\alpha$ line maps were generated by fitting the line using Gaussian functions. Maps with the \ha equivalent width and the \nii lines were also generated. In addition, maps with the oxygen  (\oiii$\lambda\lambda$4959,5007), and \hb lines for galaxies observed with INTEGRAL were produced. In this paper, we use the maps published in~\cite{Garcia-Marin09a} and~\cite{Rodriguez-Zaurin10} for the systems in our sample.

\section{Results}
\label{sec:results}

\subsection{Identification of the \ha\onespace-emitting complexes}
\label{sub:identification}

\begin{sidewaysfigure*}
\vspace{0.76\textwidth}
\hspace{0.5cm}
   \includegraphics[trim = 0cm 0cm 0cm 0cm,width=0.95\textheight]{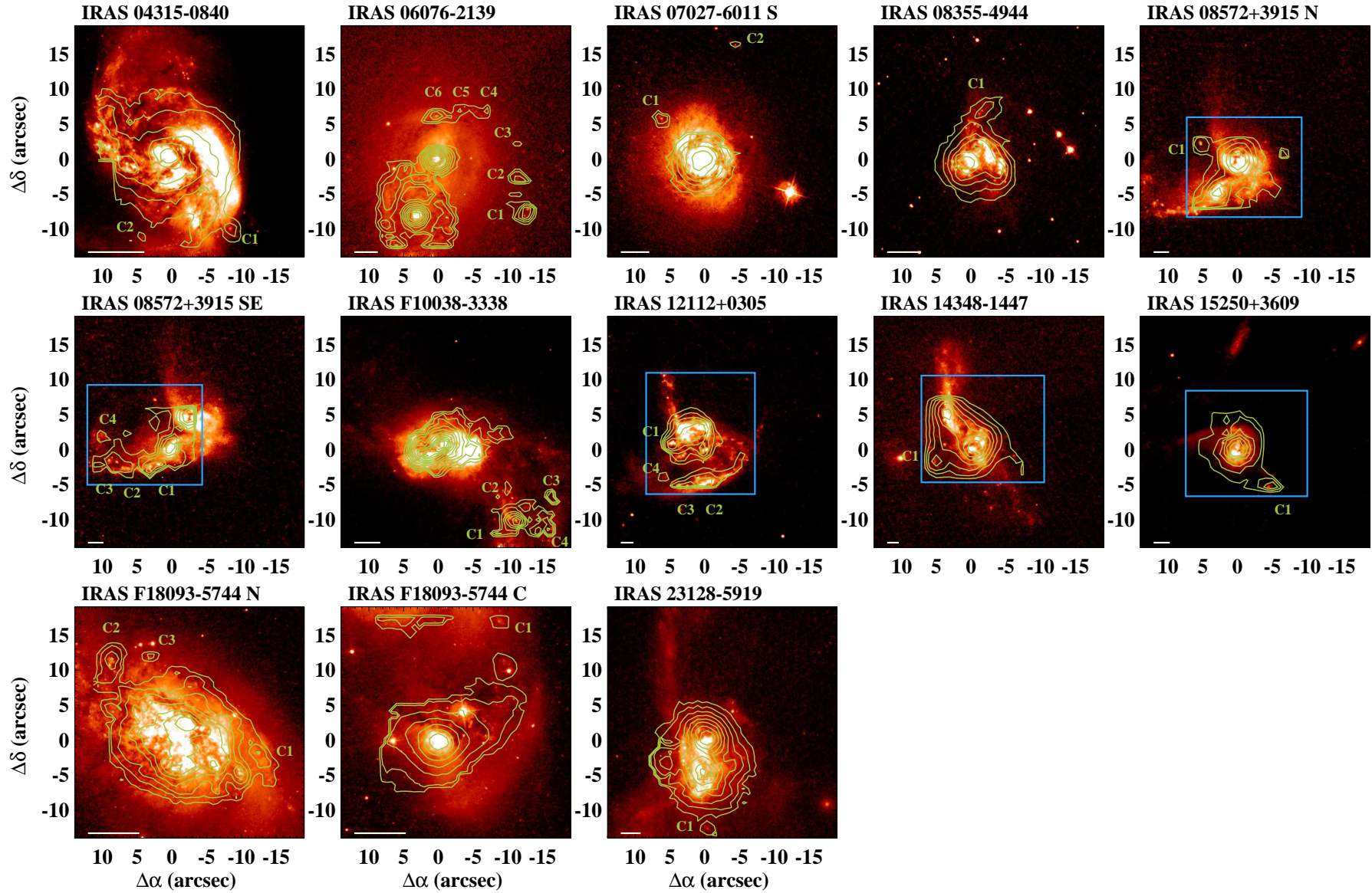} 
\caption{Systems under study with the H$\alpha$ maps obtained from integral field spectroscopy. The \textit{F435W} ACS images are shown and contours corresponding to the H$\alpha$ are over-plotted. The contours are in arbitrary units, optimized to show and identify the structures of the \ha complexes. In some cases, there are two different pointings for the same system. The labels show the location of the complexes under study. The horizontal line at the bottom left corner shows a scale of 2.5 kpc. The VIMOS FoV is shown approximately. For INTEGRAL sources, the INTEGRAL FoV is over-plotted in the blue boxes. Most of the \ha peaks match up with several knots identified in the blue images.  Note that the brightest knot on the \ha clump below C1 corresponds to a red foreground star, not a knot of star formation. \label{fig:hst_halfa} }
 \end{sidewaysfigure*}

\begin{sidewaysfigure*}
\vspace{0.75\textwidth}
\hspace{2cm}
   \includegraphics[trim = 0cm 0cm 0cm 0cm,width=0.85\textheight]{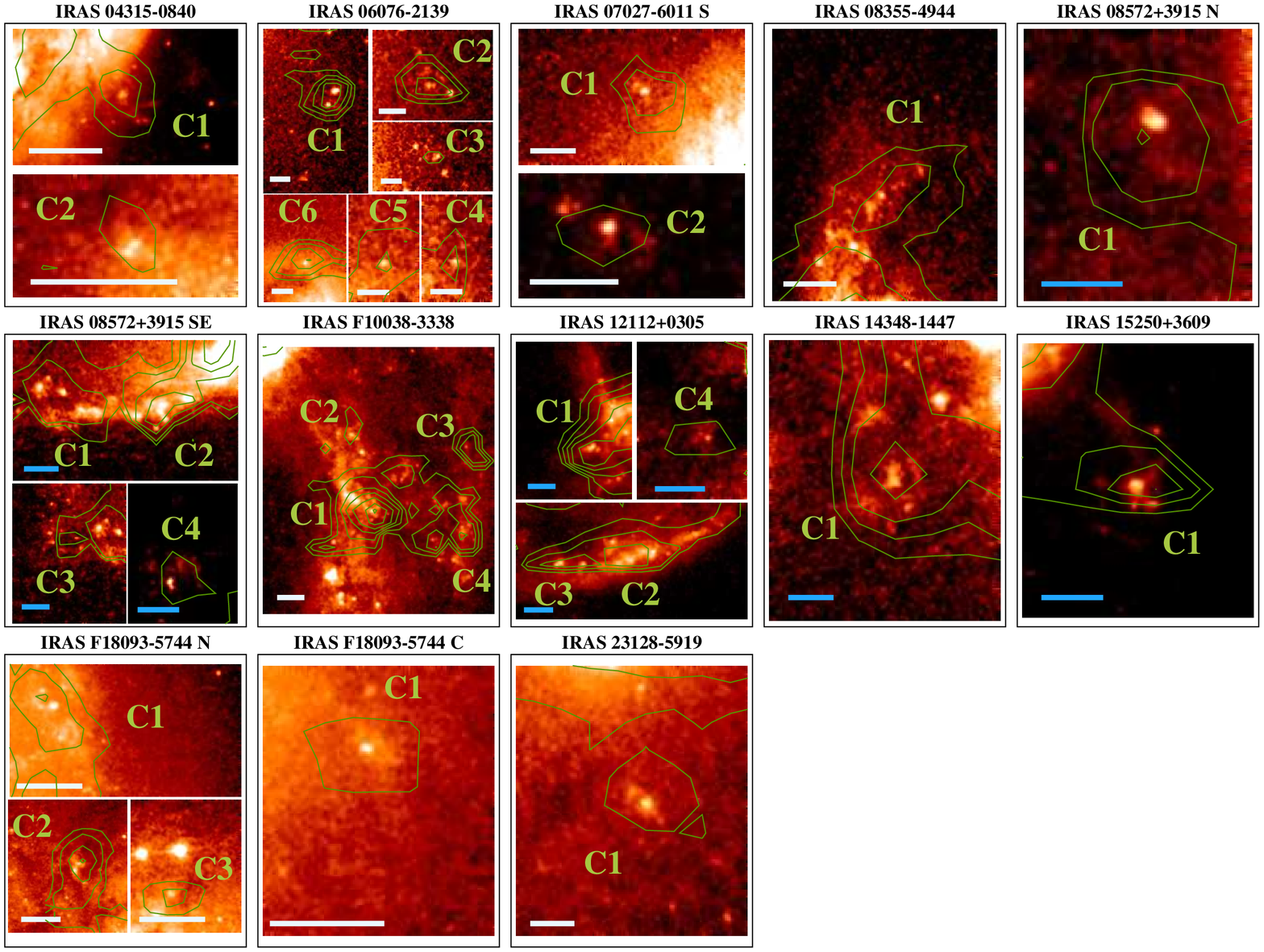} 
\caption{Zoomed view of all the identified \ha complexes. In each case, the FoV has been selected to visualize the complex more clearly. The horizontal line at the bottom left corner shows a scale of 2 kpc (blue) for complexes detected on INTEGRAL maps and 1 kpc (white) for complexes detected on VIMOS maps.\label{fig:hst_halfa_zoom} }
 \end{sidewaysfigure*}

The identification of the stellar population that ionizes the surrounding interstellar medium, and subsequently produces the \ha emission, is not necessarily straightforward. Depending on the resolution of the data and the contamination of the field by other ionizing sources, the establishment of this matching is not always possible. We first selected bright condensations in the \ha maps that were coincident with at least one blue knot in the \textit{F435W} ACS images. We chose to use these high resolution images instead of the \textit{F814W} ones because the most likely responsible for the ionization were detected most efficiently with the blue filter.  

We defined a complex of young star formation as an \ha clump (a set of spaxels\footnote{A spaxel corresponds to the spatial element from which we have one independent spectrum, and it is used to reconstruct the map.}) that had coincident position with one or several knots identified in the \hst images. In general, around 12 spaxels \mbox{(e.g., 3 $\times$ 4} spaxels) are needed to define a complex, depending on how isolated it is and the extent of the \ha emission before it is affected by another source or diffuse nuclear emission. In one case (IRAS 08572N) a prominent \ha peak is observed at around 7 kpc west of the northern nucleus, but with no corresponding region in the \hst images. 

We visually inspected each \ha map and found 11 systems with prominent peaks apart from the nuclei (see Fig.~\ref{fig:hst_halfa}). We defined the inner complexes as those located within a projected distance of 2.5 kpc to the closest nucleus and the outer complexes as those further out than 2.5 kpc. Given the resolution of the spectral maps (a factor of more than a hundred in area),
the identification of the ionizing stellar knots is highly uncertain in inner complexes, since the \ha flux there is very much likely to be contaminated by flux from nearby knots. Together with the motivation of searching for TDG candidates, they constitute the main reason why we exclude these complexes from the study. Therefore, we focus on the outer complexes. In addition, we do not select complexes in rings of star formation (i.e., those at the external ring in IRAS 06076-2139), since they are not normally associated with tidal forces, hence with TDGs. All considered, we identified for our study 31 outer young, \ha\onespace-emitting complexes of star formation, shown in Fig.~\ref{fig:hst_halfa} and~\ref{fig:hst_halfa_zoom}. 

\subsection{Structure and location of the complexes}

Owing to the higher angular resolution of the \hst images, we identified the \ha clumps (observed in the spectral maps) with a complex of star-forming knots (observed in the photometric images). The number of knots in the ACS image per \ha complex is on average 2.2. However, the structure of these clumps is in general quite simple. As can be seen in Fig.~\ref{fig:hst_halfa_zoom} and table~\ref{table:phot_prop}, the position of the \ha\twospace-emission peak is coincident or quite close to a prominent blue knot. In most cases, only one bright knot is observed within the area defined for the \ha complex (e.g., C1 in IRAS 08572+3915 N) or a centered bright knot plus a few fainter knots located at the border of the \ha complex (e.g., C1 in IRAS 04315-0840). In a few cases, the structure is quite complex, with several bright blue knots spread inside the area of the clump (e.g., C2 in IRAS 12112+0305).

When characterizing the \ha complexes, given the rich structure of some of them, we assume that every knot detected in the \hst images within the area of the complex is an \ha emitter. With the current data, the estimate of the contribution of each knot to the total \ha emission or if a knot is not an \ha emitter (i.e., it is older than 10 Myr, at which the \ha emission declines abruptly) for all the complexes, is not possible. We also assume that only the detected knots are responsible for the whole \ha emission of the complex. Throughout the paper, we analize the validity of this hypothesis, since a priori knots below our detection limit in the ACS images might also contribute to the \ha emission.

The \ha complexes are located at distances in the range 3.3-13.7 kpc from the nucleus of the parent galaxy, with a median value of 9.3 kpc (see table~\ref{table:phot_prop}). These distances, owing to projection effects, represent lower limits to the real distances. They are relatively close distances to the parent galaxy compared to those of the TDG candidates(35-100 kpc; e.g.,~\citealt{Duc98};~\citealt{Sheen09};~\citealt{Hancock09}), although some TDG candidates have been reported at such distances (e.g.,~\citealt{Iglesias-Paramo01};~\citealt{Weilbacher03}). At these distances, some of the \ha complexes lie along the tidal tails (see Fig.~\ref{fig:hst_halfa}), and the diffuse low surface brightness \ha emission (which is not subtracted in our measurements of the \ha flux) is either  undetected or considerably diminished, thus does not contaminate the emission from the knots. Others also appear to be simply a non-nuclear giant \hii region.

\subsection{Characterization of the \ha\onespace-emitting complexes}

We now investigate the physical properties of the bright \ha complexes identified. This characterization includes the analysis of the stellar continuum magnitudes (\mb and color),  spectral features (\ha luminosity and equivalent width), as well as physical sizes and an estimate of the metallicity. All these observables are compared to those measured in extragalactic \hii regions, dwarf galaxies, and other TDG candidates.

\subsubsection{Photometric properties of the complexes}
\subsubsubsection{Broad-band luminosities and colors}

The integrated blue absolute magnitude \mb and color (\mbi\twospace) of the young stellar population within the \ha complexes were estimated by adding all the flux of the knots inside a complex. We cover a magnitude range \mbox{\mb= [-9.32,-15.77]}, and a color range  \mbox{\mbi= [-0.23,2.34]} (see table~\ref{table:phot_prop}). The median values correspond to -12.06 and 0.63 magnitudes, respectively. The typical uncertainties are between 0.05 and 0.1 magnitudes. 

The integrated luminosities are in general higher than in embedded clusters in nearby extragalactic \hii regions in spirals (\citealt{Bresolin97}) and more typical of those found in extragalactic giant \hii regions (\citealt{Mayya94};~\citealt{Ferreiro08}). Although early studies in extragalactic \hii regions only focus on nearby galaxies (\mbox{\ld $<$ 20 Mpc}), Ferreiro and co-workers observed minor mergers at distances ({\ld = 40-170 Mpc}) similar to our sample, detecting also only the most luminous giant \hii regions, the external ones covering the range \mbox{$M_B~\simeq$ [-11.7, -17.4]}, typically 2 magnitudes more luminous than the range sampled in this study.

Nearby dwarf galaxies are typically more luminous than \textit{B} $\lesssim$ -13.0, with colors in the range \mbox{\mbi = 0.7-2 mag}~\citep{Hunter86,Marlowe97,Cairos01}. In addition, the luminosities observed for nearby (\citealt{Duc97,Duc98,Duc07}) and more distant (\citealt{Weilbacher00};~\citealt{Temporin03}) TDG candidates are generally more luminous than \mbox{\textit{B} = -10.65}, as for most of our \ha complexes.

The integrated broad-band luminosities of the embedded stellar population in our \ha complexes is, therefore, compatible with the luminosities measured in giant \hii regions and TDG candidates.

\subsubsubsection{Sizes and compactnesses}

Owing to the higher angular resolution of the \hst images, estimates of the sizes and compactnesses of the \ha emitting clumps were derived from the ACS images. All the knots within a complex were assumed to represent the ionizing stellar population. We then added up the area of each knot (derived in MC11) within an \ha complex, computing a total area $A_{\rm{T}}$. With this area, we derived an equivalent total radius, given by $r$=$\sqrt{A_{\rm{T}}/\pi}$ (see table~\ref{table:phot_prop}). Had we used the spectral maps to measure the radius of the complexes ($r_{\rm{H} \alpha}$ in table~\ref{table:phot_prop}), we would have overestimated it in general by more than a factor of two. The radii $r$ range from somewhat less than 100 pc to about 900 pc, with a median value of 280 pc. These sizes are similar to those for the largest giant extragalactic \hii regions observed in nearby (\mbox{\ld $<$ 20 Mpc}) spirals (\citealt{Mayya94};~\citealt{Rozas06};~\citealt{Diaz07}), measured using ground-based instrumentation.

To evaluate the compactness of the \ha complexes, we also derived an effective radius (\reff\twospace) for each, as an approximation of the half-light radius. We again used the \hst images for this determination. In this case, since in most cases one knot dominates the total broad-band luminosity within the complex, we identified the effective radius of an \ha complex  with the effective radius of the most luminous knot (derived in MC11). In complexes where more than one knot dominates the luminosity, this approach lead to an underestimate of the effective radius.

The effective radii of complexes observed with VIMOS are smaller than 100 pc (see table~\ref{table:phot_prop}), from 20 pc to about 80 pc. For those observed with INTEGRAL, the values range from  100 pc to about 300 pc. These sizes are in general larger than the effective radii of the so-called ultra compact dwarf galaxies (e.g.,~\citealt{Dabringhausen08}), similar to those in giant \hii regions (e.g.,\citealt{Relanyo05,Rozas06}) and within the range of  the smallest blue compact dwarf galaxies (with \mbox{\reff=0.2-1.8 kpc};e.g.,~\citealt{Cairos03,Papaderos06,Amorin09}) and Local, NGC 1407, and Leo groups (with \mbox{\reff $\sim$ 0.3-1} kpc;~\citealt{Mateo98,Forbes11}). 

The ratio of the effective to equivalent radii gives an idea of the compactness of a certain complex. They are very compact, with a ratio of typically \mbox{\reff\twospace/$r$ = 0.2}, ranging from 0.15 to 0.6.

\subsubsection{\ha luminosities and equivalent widths}

\begin{figure}
\hspace{-0.3cm}
\includegraphics[angle=90,width=1.\columnwidth]{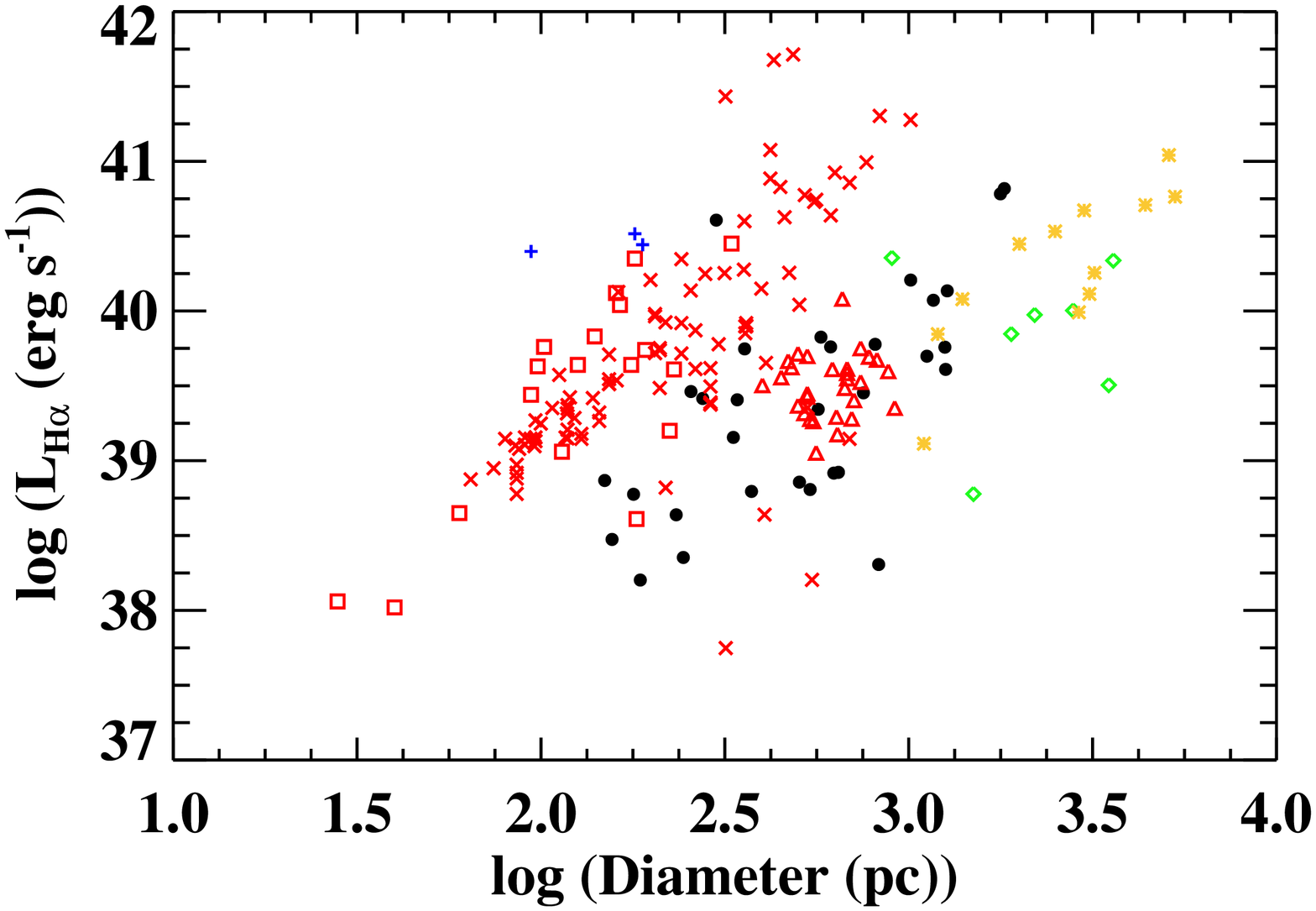}
\caption{Relation between the \ha luminosity and the size of the identified \ha complexes in our sample (black dots), extremely luminous \ha complexes in the Antennae (blue plusses;~\citealt{Bastian06}), different samples of isolated giant \hii regions in spirals and minor mergers (red symbols;~\citealt{Terlevich81,Ferreiro04,Rozas06}), TDG candidates in CG J1720-67.8 (green diamonds;~\citealt{Temporin03}) and nearby dwarf amorphous (yellow crosses;~\citealt{Marlowe97}). None on the  luminosities on the plot are corrected for internal extinction. \label{fig:halfa_lit}}
\end{figure}

Both \ha flux and equivalent width measurements (EW) were obtained for the 31 \ha complexes. Their observed \ha luminosity is typically higher than \mbox{10$^{39}$ \ergs} (see table~\ref{table:spec_prop}). The luminosity of about 20\% of them is even higher than \mbox{10$^{40}$ \ergs\twospace}. The median value of the \ha flux is \mbox{2.9$\times$10$^{39}$ \ergs\twospace}.~\cite{Monreal07} also measured the \ha fluxes for four of our systems with INTEGRAL data (in IRAS 08572+3915 only for the northern pointing) and most of them are systematically lower by up to a half. The discrepancy occurs because they used apertures of 0.5\arcsec or 1.0\arcsec, whereas we have added up the fluxes of several spaxels, considering a greater effective aperture (one spaxel on INTEGRAL maps already corresponds to an aperture of 0.45\arcsec~in radius). In the case of IRAS 12112+0305, we grouped their kc and k1 regions and did not consider R2 because it is too close to the northern nucleus, which makes its \ha flux determination very uncertain.

Fig.~\ref{fig:halfa_lit} compares the \ha luminosities and sizes of our complexes with those of extragalactic \hii regions, \ha complexes in nearby galaxies, TDG candidates, and dwarf galaxies. This plot shows that using the \ha luminosity alone to establish the nature of a given object might be misleading. For instance, the complexes of star formation in the Antennae have an \ha luminosity comparable to the brightest extragalactic \hii or even to dwarf galaxies, while these complexes are generally much smaller. Our complexes are typically located where the giant extragalactic \hii regions and the dwarf-type objects lie on the plot.

\begin{figure}[t!]
\includegraphics[angle=90,width=1.\columnwidth]{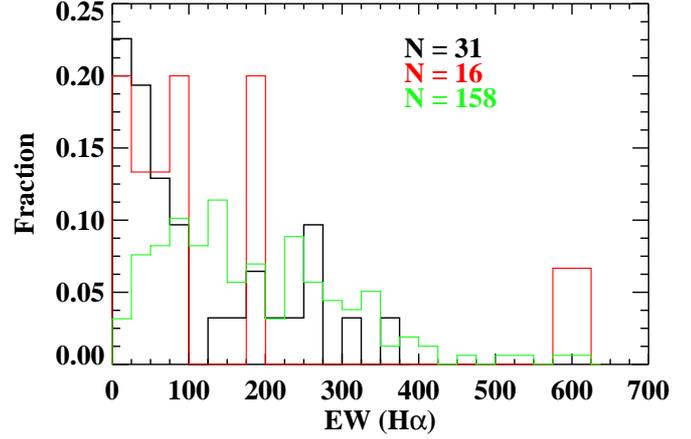}
\centering
\caption{Equivalent widths measured for the complexes  in this study (black), for extragalactic \hii regions (green;~\citealt{Mayya94}) and for TDG candidates (red;~\citealt{Iglesias-Paramo01,Temporin03}).The number of objects is indicated in each case. \label{fig:ew}}
\end{figure}

We have also estimated the total equivalent width of a given complex from the EW and the \ha spectroscopic line maps. Since in a spaxel the EW $\propto$ {\small{Flux}} \ha$\!\!$ / {\small{Flux continuum}}, we first determined the continuum per spaxel. We then added up all the continuum within a complex to compute the total continuum of the complex. Finally, we divided the total \ha flux by the total continuum flux for each complex. Their integrated EW span a range of two orders of magnitudes, from 4.1$\AA{}$ to about 170$\AA{}$ (see table~\ref{table:spec_prop}). For each complex, we also give the value of the spaxel with the highest EW (EW$_{\rm{peak}}$). The highest value corresponds to EW$_{\rm{peak}}$=374$\AA{}$. Its median value (\mbox{$<\rm{EW}_{\rm{peak}}>$ = 85$\AA{}$}) doubles the median value of the integrated EW computed for the complexes. The \ha equivalent widths derived, which is indicative of a very young stellar population, are comparable to those measured for extragalactic \hii regions and  TDG candidates  (see Fig.~\ref{fig:ew}). 

The ratio $\frac{\rm{Peak}~\rm{EW}}{\rm{EW}~(\rm{H}\alpha)}$ was defined for each complex, where \mbox{\textit{Peak EW}} refers to EW$_{\rm{peak}}$. This ratio helps us understand how the ionizing population is distributed. If two complexes have a similar underlying old population, the ionizing population in the one with the larger ratio is probably more concentrated (i.e., less knots) than the other with the smaller ratio. And if it is less concentrated (i.e., has more knots),  most of the knots are probably somewhat older (low EW) and only a few are very young (high EW), leading to an older average population. In fact, the complexes with large ($\gtrsim$ 3) ratios have typically few knots (1-3), such as C1 in IRAS 06076-2139, C1 in IRAS 08572+3915 N, and C4 in IRAS 12112+0305.

\subsubsection{Contamination of the \ha luminosity from embedded young populations}

With the knowledge of the integrated broad-band luminosities, the \ha luminosities, and the equivalent widths of the complexes, we can try to evaluate whether the measured \ha flux is emitted by possibly embedded (and undetected) star-forming knots. The existence of undetected \ha\twospace-emitters is justified with the \ha\twospace-emitter in IRAS 08572+3915 N (clump in the north-west).

We estimated how many undetected knots there can be with an age derived from the peak values of the EW (typically between 5 and 10 Myr, according to the stellar population models used in section~\ref{sub:age_mass_complex}) that can contribute significantly to the total broad-band luminosity of the knots. Since they would have a similar age, an embedded population of knots with a combined broad-band flux similar to that of the detected knots would emit half of the \ha flux. We then estimated how many knots at the detection limit would be needed to double either the \textit{F435W} or the \textit{F814W} flux. A total of  9 and 75 undetected knots per complex would be needed and we detected an average of 2.2 knots per complex. Thus, between 4 and 37 times more knots fainter by 2.4-4.7 magnitudes would be needed only in one complex. Assuming a luminosity function with slope of 2 we would need practically one-third of the total knots predicted for the whole galaxy inside a complex, which is unlikely.

\subsubsection{Metallicities}

\begin{figure}
\hspace{-0.5cm}\includegraphics[angle=90,width=1.05\columnwidth]{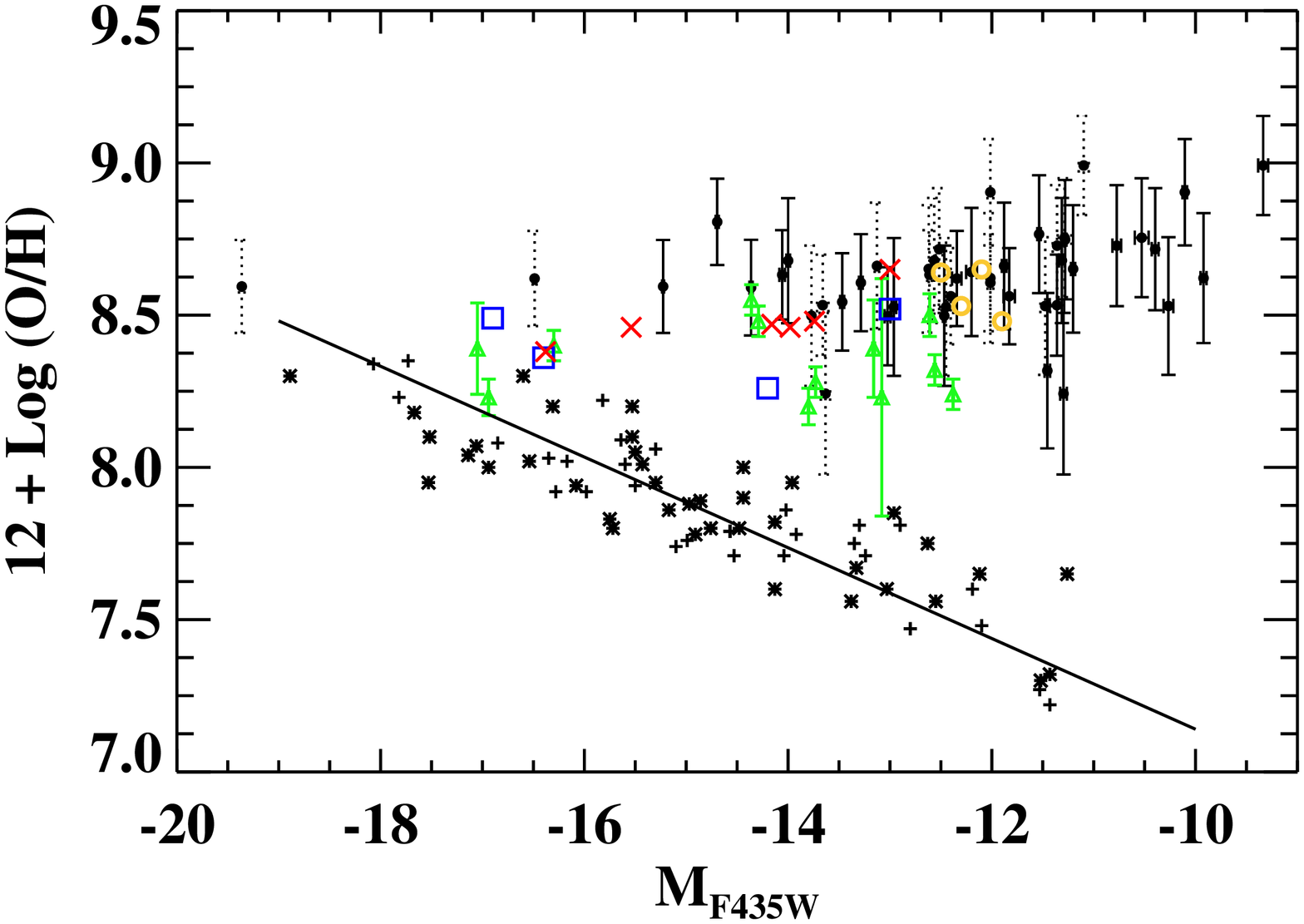}
\caption{Metallicity-luminosity relation for external \hii regions in Stephan'{}s Quintet (yellow open circles;~\citealt{Oliveira04}), nearby irregular galaxies (black plusses,~\citealt{Pilyugin04}; black asterisks,~\citealt{VanZee06}), TDG candidates (green filled triangles,~\citealt{Weilbacher00,Weilbacher03}; red crosses,~\citealt{Duc98}; blue open squares,~\citealt{Temporin03}) and our complexes (black dots with errors). The line shows the correlation found by van Zee. Only the luminosities given by Pilyugin are corrected from internal extinction. For those \ha complexes for which an estimate of the extinction has been made (see section~\ref{sub:age_mass_complex}), the metallicity-luminosity relation is also shown (dots with dashed line errors).\label{fig:metal_mb}}
\end{figure}

Numerical studies have claimed that during an interaction large amounts of gas flow toward the central regions, carrying less enriched gas from the outskirts of the galaxy into the central regions (\citealt{Rupke10};~\citealt{Montuori10}). Indications of this mixing process, which usually disrupts any metallicity gradient and dilutes the central metallicity, have been observed in merging (U)LIRGs (\citealt{Rupke08}), in galaxy pairs (\citealt{Kewley06};~\citealt{Ellison08};~\citealt{Michel-Dansac08}), and in star-forming galaxies at high-z (\citealt{Cresci10}). As a consequence of the metal enrichment of the external regions of the mergers, additional star formation can occur from reprocessed material there. We investigated whether the interaction process affects the metallicity of our \ha complexes.

Abundances are usually estimated using empirical methods based on the intensities of several optical lines. The most popular methods are the widely used R23 (\citealt{Torres-Peimbert89}) and S23 (\citealt{Vilchez96}) calibrators. However, both methods use emission lines that our spectral range does not cover. We used instead the N2 calibrator proposed by~\cite{Denicolo02} and the empirical diagrams of~\cite{Edmunds84}. The former is based on the ratio of the \nii\twospace$\lambda$6584 to the \ha emission lines and the latter relates the ratio \oiii\twospace$\lambda$4959 + \oiii\twospace$\lambda$5007 to \hb to the oxygen abundance, parametrized as in \cite{Duc98}. 

Given the spectral range of the VIMOS data, metallicity determinations have only been possible using the N2 calibrator for systems observed with this instrument. The typical uncertainties in the determined metallicities are about 0.2 dex. For the candidates observed with INTEGRAL, given these uncertainties, the differences between both indicators (usually within 0.2 dex) are irrelevant, thus we adopted the average value. The adopted metallicities for all the candidates are shown in the last column in table~\ref{table:spec_prop}. With values compatible to a solar metallicity, they generally range from 12 + logO/H = 8.5 to 8.8 (with the exception of two candidates with lower abundances). The values for complexes observed with VIMOS have to be considered with care and must be checked with other metallicity indicators, since the N2 calibrator might be affected by ionization from strong shocks in the external regions of these galaxies (\citealt{Monreal10}). In addition, spatially resolved studies of star-forming regions indicate that the assumption of spherical geometry is unrealistic in most cases, which has a direct impact on the derivation of metallicities.~\cite{Ercolano07} estimated the systematic errors in the metallicity determinations when assuming spherical geometry using different calibrations. In the worst-case scenario, the derived oxygen abundances might be overestimated by 0.2-0.3 dex by the use of the calibrators considered in this paper, in which case the derived metallicites for the \ha complexes would then approach Z$_{\odot}$/3.

The \ha complexes in this study do not follow the well-known metallicity-luminosity relation for nearby isolated dwarf galaxies (see Fig.~\ref{fig:metal_mb}). Most galaxies also follow this relation (\citealt{Weilbacher03}). On the other hand, \hii regions in compact groups of galaxies and TDGs in general deviate significantly from the relation and have a metallicity that is independent of their luminosity, an indication that all these objects consist of recycled material.

\section{Discussion} 
\label{sec:discussion}

Most \ha\twospace-emitting complexes have similar observational properties (i.e., \ha equivalent widths and luminosities, \mb magnitudes, metallicities, radii) to the most luminous extragalactic giant \hii regions in spirals and mergers, as well as more massive objects such as \tdgs or \tdg progenitors. Associations of young star-forming regions with a large HI reservoir have been found in TDGs, for instance by~\cite{Duc07}. We may then consider whether the \ha\twospace-emitting complexes in our sample are dynamically unbound associations of objects with masses similar to observed super star clusters (SSCs) or larger, more massive, and self-gravitating objects such as dwarf galaxies.

\tdgs are self-gravitating objects with masses and sizes typical of dwarf galaxies (i.e., a total mass of 10$^7$-10$^9$~\msun\twospace), formed with recycled material from the parent galaxies involved in an interaction/merger (\citealt{Duc00}). Hence, they are stable entities with their own established dynamical structure. To evaluate whether our complexes constitute real TDGs or TDG progenitors and assess their chance of survival, we need to answer a few basic questions: Is the complex massive enough to be considered as a dwarf galaxy? Is it stable enough to be unaffected by its internal motion? Are the gravitational forces too strong to disrupt it? In this section, we disscuss these questions.

\subsection{Selection of \ha\twospace-emitting likely to represent TDG candidates}

Although old \tdgs have been observed (e.g.,~\citealt{Duc07}), their identification is not straightforward. When an old \tdg is detected, the tail from its place of origin will probably have completely disappeared and the TDG may be classified as another type of dwarf galaxy. A determination of the metallicity and, especially, the total mass of the objects is needed to establish its tidal origin and mass. With spectroscopic data, we are able to study (though with some limitations and biases) the metal content, the dynamical mass, and different methods to establish whether a given candidate can withstand both internal and external forces. It is quite complicated to identify luminous enough condensations of old populations (i.e., very massive) and obtain kinematic information because the stellar absorption lines that would have to be used often have too low a signal-to-noise ratio (S/N). Although \tdgs containing only old populations may be present in our sample, we are unable to detect them. Another way of searching for \tdg candidates is based on either analyzing the \ha emission clumps (for that we have a sufficiently high S/N) or/and combining spectral and photometric data of blue \ha\twospace-emitting objects in interacting systems (e.g.,~\citealt{Iglesias-Paramo01,Weilbacher03,Temporin03,Lopez-Sanchez04}). Therefore, the selection of \tdg candidates in our sample of (U)LIRGs is based on the detection of \ha emission clumps with stellar counterparts, which is similar to our identification of \ha complexes.

However, we are limited by the angular resolution of the spectral data. To make our estimates as accurate as possible, we do not consider in our discussion here complexes with multiple luminous star-forming knots. Different types of motions within the components of a given complex can strongly affect the derivations mentioned before. The most characteristic examples of complicated complexes are C1 in IRAS F10038-3338 and C2 in IRAS 12112+0305. This does not mean that the possibility of these complexes being \tdgs is ruled out, just that we  do not have enough resolution to study the kinematics of each knot, thus determine whether they are kinematically bound or not. 

We consider only complexes with simple structures, that is, mainly where only one knot is detected. We also include complexes with several knots, one of which (normally centered on the \ha peak emission) dominates the broad-band luminosity. The complex C1 in IRAS 04315-0840 would be a good example. On the basis of these criteria, the kinematic and dynamical properties of 22 complexes are derived and compared with those expected for a \tdg\twospace.

If we detect more than one knot within the selected complexes, we consider only the broad-band luminosity of the brightest knot. In these cases, the \ha luminosity of the complex is scaled in the same way as the \textit{F435W} flux is scaled to the total \textit{F435W} broad-band flux of the complex in table~\ref{table:phot_prop}. In practice, we multiply the \ha luminosity by a factor that in general is higher than 0.5 (a factor of 1 means that there is only one knot emitting the whole broad-band flux, thus responsible for the whole \ha emission). The kinematic measurements, such as the velocity dispersion, are not performed for the whole complex, but only for the spaxels close to and located at the peak of the \ha emission. This permits us to avoid any contamination by a faint knot in the border of the complex (e.g., faint knots within C1 in IRAS 04315-0840).

Using the kinematic information provided by the IFU data as well as the photometric measurements, we assess the origin and likelihood of survival of the 22 selected \ha complexes, and evaluate whether they can be the progenitors of tidal dwarf galaxies.

\subsection{Are the selected \ha complexes massive enough to constitute a TDG?}

\subsubsection{Age and mass estimates of the young population}
\label{sub:age_mass_complex}
 
The \ha emission can be used to constrain the properties of any recent episodes of star formation. Given the youth of the population we study, we use the single stellar population synthesis model Starburst99 (SB99,~\citealt{Leitherer99,Vazquez05}), which have been optimized for the analysis of young populations, to estimate its age and mass. We consider instantaneous burst models with a Kroupa IMF (\citealt{Kroupa02}) over the range 0.1-120~\msun and solar metallicity. These models are normalized to 10$^6$~\msun\onespace. Mass estimates are multiplied by a factor of 1.56 when the Salpeter IMF is considered.

According to the SB99 models, the detection of an \ha flux and EW larger than about 13 \AA{} (see table~\ref{table:spec_prop} for specific values) implies that the stellar population is younger than 10 Myr. With the broad-band luminosities of the brightest knot within the complex and the scaled \ha luminosity, we can estimate to first order the age and the mass of the young population of the selected complexes. We note that, owing to the distance of the systems (\mbox{\ld$<$ 270} Mpc) and the shape of the \hst filters, the \ha emission-line does not contaminate these broad-band filters. 

\begin{figure}
\includegraphics[angle=90,width=1.05\columnwidth]{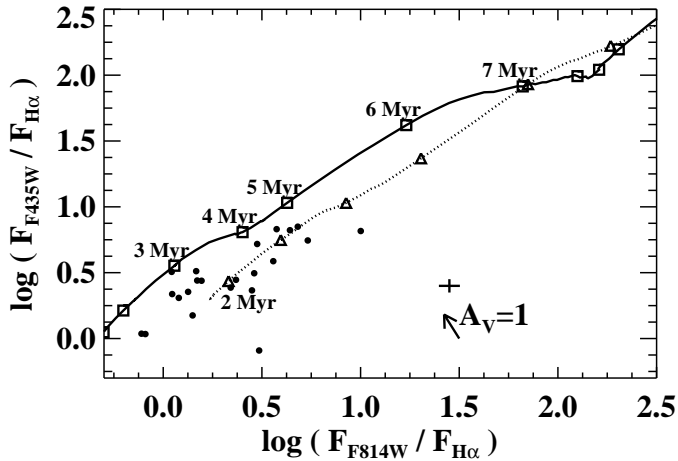}
\caption{Flux ratios of the blue and red filters when the \ha emission can be measured in a complex. The broad-band filter fluxes have been multiplied by the band-width to obtain the values with the same units as the \ha flux. Solid line shows the evolutionary track of the same ratios for a single burst using SB99 models (see text). Dotted line shows the track for a composite population where 99\% of the mass corresponds to a 1Gyr population. The small triangle almost at the beginning (zero-age young population) of the second track indicates that the young population is 2Myr old and we add 1 Myr to the subsequent triangles. The typical uncertainty associated with the data is shown above the extinction vector, which is drawn at the bottom right-hand corner. \label{fig:halfa_com}}

\end{figure}

We computed the mass-independent evolutionary track shown in Fig.~\ref{fig:halfa_com} using information about the ratio of the flux of the modeled population in the broad-band filters to the flux of the ionized gas in the \ha line. Placing these ratios on a plot for all observed complexes, we now estimate the age of the young population. Once the age is known (hence a M/L ratio), the mass can be derived. However, before going any further and based on the assumptions made, we first assess the sources of uncertainty that a priori could affect our results:

\begin{itemize}

\item  The internal extinction is unknown. In general, the extinction in the inner regions is patchy with high peaks in (U)LIRGs, up to \av $\simeq$ 8 mags (see~\citealt{Alonso-Herrero06} and~\citealt{Garcia-Marin09b}), though it decreases considerably with the galactocentric distance. Were complexes to have low extinction values (\av $\lesssim$ 1.5mag) their flux quotients would tend to follow the evolutionary track (solid line in Fig.~\ref{fig:halfa_com}), and the age would remain practically unchanged. Only one complex probably has a higher internal extinction (C1 in IRAS 14349-1447). A high extinction value of \av $\sim$ 3.5 mags was indeed measured for this complex in~\cite{Monreal07}. The typical extinctions of the TDG candidates studied in Monreal-Ibero are between \mbox{\av= 0.7 and 2 mag}, which is consistent with the extinctions derived for our selected complexes (see Fig.~\ref{fig:halfa_com}).\\

\item The local background flux (from the underlying parent galaxy) assumed to be associated with either older populations or non-\ha line-emitters, was subtracted when the photometry was performed for the knots. However, a knot itself can be formed by a composite of young and old population. If we were to assume that a significant fraction of the red and blue fluxes measured for the knots originated from an older population, the evolutionary track would change (see the composite populations in Fig.~\ref{fig:halfa_com}); the young population would then be even younger than initially predicted and consequently less massive. Almost half of the complexes are incompatible with the composite track, since their flux ratios would be too small, even smaller than for a zero-age young population. The other complexes, for which the values are compatible with the composite track, would typically have a young population of about or younger than 4 Myr. Were this to be the case, the extinction would be negligible and the equivalent widths would be about 200 \AA{}. \\

\item Had we included some \ha contribution from the neighboring zones then the \ha flux measured would have been overestimated. The \ha flux drops considerably as age increases across the age interval we consider (by up to more than one order of magnitude), but if the \ha flux is weaker the mass of the young population decreases accordingly. The mass estimated using the red filter can vary by up to a factor of four within the age interval 1-10 Myr, according to the stellar population models used in this study. The age estimate would be older by no more than 1 Myr, if we had overestimated the flux by a half. Thus, this correction would in our case be negligible.

\end{itemize}

Taking into account these sources of uncertainty, the first-order estimates of the age were performed for the young population inside a complex, assuming a single burst. We estimated the age by considering the extinction value that we would need to shift the location of each dot (flux quotients) along the evolutionary tracks. The estimated ages range from 2 to 5 Myr (Age$_{phot}$; see table~\ref{table:derived_prop}), the typical age range in which an instantaneous burst of star formation displays Wolf-Rayet (WR) features in its spectra (\citealt{Leitherer99}). Given the spectral range covered in this study, we would expect to see the well-known red WR bump, which is the result of the blend of the \ciii $\lambda 5698$ and  \civ $\lambda 5808$ broad emission lines (\citealt{Kunth86}). We tried to find these features in the spectra of the complexes, but could find no clear evidence of these bumps. This bump is much more difficult to detect than the other bump (such as the blue bump) characteristic of WRs because it is always weaker (e.g.,~\citealt{Fernandes04};~\citealt{Lopez-Sanchez09}). Sometimes it is not even detected again in Wolf-Rayet galaxies where it was previously seen (\citealt{Lopez-Sanchez09}). Therefore, the non-detection of WR features does not definitely exclude the hypothesis that the young population in our complexes spans the age range 2-5 Myr.

An upper limit to the age can be estimated by using the EW (\ha\twospace), as it decreases strongly with time. Using the EW peaks (see table~\ref{table:spec_prop}), we estimate that the age is in the range \mbox{5-10} Myr (Age$_{EW}$; see table~\ref{table:derived_prop}). In many cases, the EW peaks are not strong enough to explain population as young as our former age estimates. This is unsurprising because within a complex and a spaxel itself the underlying older population contributes significantly to the \ha continuum but not to the emission line. The contribution of this continuum minimizes the total EW within a complex. The broad-band flux of the knots typically represents between 1\% and 40\% of the total flux within the area of the complex. We consider for instance a 3 Myr-old population whose broad-band flux represents 10\% of its overall broad-band flux within the complex. According to the models used in this paper, the equivalent width of the young population without contamination would be $\sim$1000 \AA{}. However, the contamination of a 1 Gyr-old population would diminish the equivalent width to a measured value of 70 \AA{} for the whole complex, the same value as a single population of 6 Myr without contamination. Therefore, with the observed EW we can only set an upper limit to the age in each case.

As a conservative approach, instead of directly using the youngest estimates to derive the mass of the young population, we use the average of the two estimates, Age$_{phot}$ and Age$_{EW}$. Thus, once the age and extinction were estimated, the mass was directly obtained via the extinction-corrected \textit{F814W} magnitude.
 
Under these assumptions, the derived mass of the young population of the selected \ha complexes is between 10$^{4.5}$~\msun and 10$^{5.5}$~\msun\twospace, with the exception of three complexes for which the derived mass is about or close to 10$^{7}$~\msun\twospace (M$_{[I]}$; table~\ref{table:derived_prop}). These complexes have extinctions of \mbox{\av $\sim$ 1-2} mag, with the exception of C1 in IRAS 14348-1447 (\mbox{\av = 4.2 mag}). The uncertainty in this mass is typically smaller than a factor of two. This rather small uncertainty is expected, since during the first 1-7 Myr of the starburst the broad-band luminosities do not change significantly. However, if we compute the mass using the \ha luminosity (M$_{[\rm{H}\alpha]}$; table~\ref{table:derived_prop}) the uncertainties increase considerably. This is also expected because the \ha flux for a population of a given mass evolves significantly with the age of the starburst. The few cases in which both measurements are incompatible suggest that the age of the burst is closer to the youngest value (Age$_{\rm{phot}}$). The corresponding M$_{[\rm{H}\alpha]}$ would be lower toward a value similar to M$_{[I]}$.

An old population of a few Gyr (prior to the interaction, and in the parent galaxy) in \tdg candidates that contributes to most of the stellar mass has been reported (e.g.,\citealt{Sheen09}). Large \hi reservoirs have also been found in a few \tdgs (\citealt{Duc07}) that can sustain star formation on a Gyr scale. We have seen previously that the colors of some complexes can be compatible with a composite population where a 1 Gyr-old population is 99\% more massive than a young burst. Thus, we consider it worth investigating this possibility also here.  Table~\ref{table:derived_prop} shows the estimates of the age of the young (Age$_{\rm{Cyoung}}$) and the mass of both populations (M$_{\rm{Cyoung}}$ and M$_{\rm{Cold}}$). The stellar mass of the complexes generally ranges from 10$^{5.5}$~\msun to 10$^{6.5}$~\msun, an order of magnitude higher than the previous estimates.

\subsubsection{Conditions for self-gravitation}
\label{sub:internal_motions}

\begin{figure}
\hspace{-0.5cm}\includegraphics[angle=90,width=1.05\columnwidth]{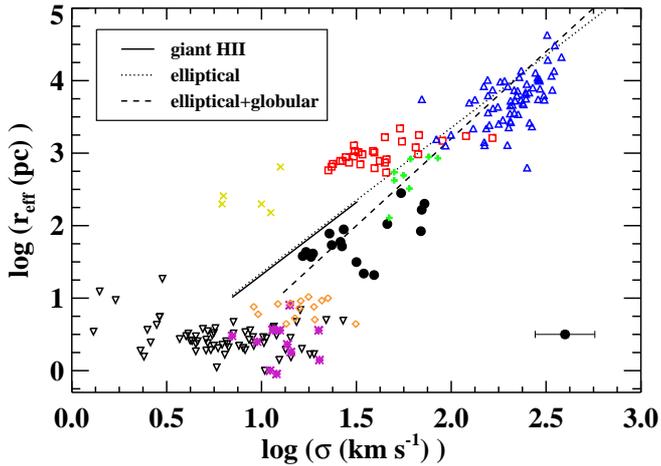}
\caption{Velocity dispersion vs. estimated effective radius. Dots correspond to the relation for our selected complexes. The typical size for errors is shown in the bottom right corner. The lines show the fit for extragalactic \hii regions (continuous), elliptical galaxies (dashed) and globular + elliptical galaxies (dotted) obtained by~\cite{Terlevich81}. The other symbols represent different samples of dynamically hot systems: open triangles in blue, intermediate and giant ellipticals (\citealt{Bender92}); open squares in red, dwarf ellipticals (\citealt{Bender92};~\citealt{Geha03}); crosses in yellow, dwarf spheroidals (\citealt{Bender92}); plusses in green, the TDG candidates in~\cite{Monreal07}; diamonds in orange, massive globular clusters in NGC 5128 (\citealt{Martini04}); inverse triangles, globular clusters in the Galaxy (\citealt{Trager93}) and in M31 (\citealt{Dubath97}); and asterisks in pink, clusters in the Antennae (\citealt{Mengel08}). \label{fig:sigma_reff}}
\end{figure}

Measuring velocity gradients helps us assess whether there is either any independent rotation (\citealt{Weilbacher02};~\citealt{Bournaud04,Bournaud08a}) or outflows. However, given the spatial resolution achieved with the IFS instruments, we cannot in general resolve any velocity field across the extranuclear condensations. In this case, other indirect methods must be used to assess whether the internal motions of our complexes can affect their stability.

\cite{Iglesias-Paramo01} established an empirical luminosity criterion (L(H$\alpha$) $>$ 10$^{39}$ erg s$^{-1}$) that must be reached by their \hii complexes to ensure self-gravitation. This criterion, however, has not been supported by any kinematic study. More than 50\% of our own complexes fulfill this criterion, as shown in table~\ref{table:spec_prop}. If we corrected for internal extinction, only three would be less luminous than 10$^{39}$ erg s$^{-1}$. 

\begin{figure}[!t]
\hspace{-0.5cm}\includegraphics[angle=90,width=1.05\columnwidth]{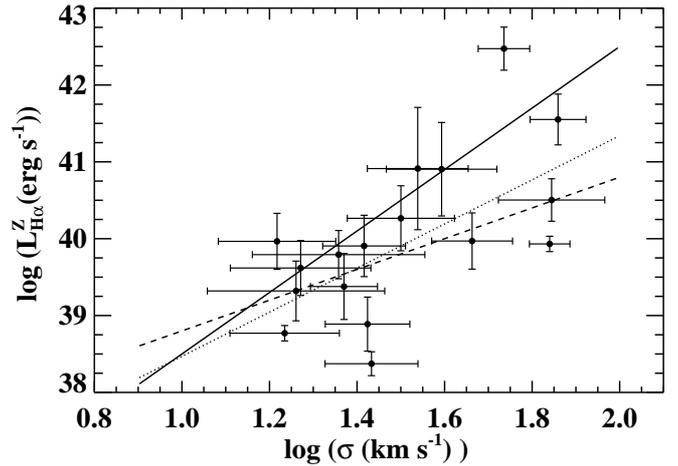}
\caption{Velocity dispersion vs. \ha luminosity corrected for internal extinction and metallicity effects. The lines show different fits for extragalactic giant \hii regions which are approximately in virial equilibrium, given by: \cite{Terlevich81} (solid line),~\cite{Relanyo05}  (dotted line) and~\cite{Rozas06} (dashed line).   \label{fig:sigma_halfa}}
\end{figure}

Another method that can be used to establish whether the selected complexes are stable is to study their location in well-known empirical correlations followed by other self-gravitating entities such as elliptical galaxies, bulges of spiral galaxies, globular clusters, and/or giant \hii regions: the radius-velocity dispersion and the luminosity-velocity dispersion relations (\citealt{Terlevich81}). Here, we consider these relations because they provide more reliable constraints than the \ha luminosity of the complexes. We plot these relations for our complexes in Figs.~\ref{fig:sigma_reff} and~\ref{fig:sigma_halfa}. 

In the radius-velocity dispersion diagram, we superimpose the data of samples of dynamically hot systems such as massive globular clusters, dwarf elliptical and spheroidal galaxies, intermediate and giant ellipticals, and \tdg candidates. Within all of the uncertainties, the velocity dispersions of six of the complexes selected here are too high to ensure self-gravitation.

For a giant \hii region, the lower envelope in the log \lha - log$\sigma$ plane is closely represented by a straight line of the form \lha = c  + $\Gamma \times log{\sigma}$, where $\sigma$ refers to the velocity dispersion, the constant c ranges from 34.5 to 36.8, and $\Gamma$ is between 2 and 4 (see line fits and references in Fig.~\ref{fig:sigma_halfa}). This is known as the luminosity-velocity (L-$\sigma$) dispersion relation, and implies that the giant extragalactic \hii regions above the envelope are gravitationally bound complexes of stars and gas and that the widths of the nebular emission lines reflect the motions of discrete ionized gas clouds in the gravitational field of the underlying stellar and gaseous mass (\citealt{Terlevich81}). The kinematics of the \hii regions below the envelope may be dominated by processes other than gravitational motions, such as stellar winds and expanding shells (\citealt{Relanyo05,Relanyo05b,Rozas06,Rozas06b}). That is, if the relation for giant \hii regions lies above the envolope, then the non-gravitational processes are identified with broad, low-intensity components that do not considerably affect the physical properties of the main spectral component (i.e., gravitational motions). Thus, although the velocity of these fenomena could be similar or even larger than the $\sigma$ measured in this paper, if the relation for our complexes is above the envelope, the spectral component corresponding to the gravitational motions will probably not be affected. 

In any case, we estimated the expansion speeds of bubbles in the interstellar medium, so as to quantify how broad these non-gravitational components are. We evaluated the equivalent number of O3(V) stars using the \ha luminosity of a given complex and the value given by \cite{Vacca96} for the ionizing photon output of an O3(V) star. We then derived the kinetic energy input from the stellar winds, using the estimate by \cite{Leitherer98} of the wind luminosity for an O3(V) star, integrated over the time the star is on the main sequence. Assuming a typical average electron density of \mbox{100 cm$^{-3}$}, obseved in (U)LIRGs (\citealt{Veilleux95,Veilleux99}), the expansion speeds are expected to be (\citealt{Lamers99}) between several km s$^{-1}$ for the least luminous complexes and less than 30 km s$^{-1}$ for the most luminous ones.

Given the metallicity range of our selected complexes, we applied a metallicity correction to the \ha luminosity, in such a way that \mbox{$\Delta$log (F(\ha\twospace))=127z - 1.17} (\citealt{Terlevich81}), where z denotes the metallicity value (z=0.02 for solar metallicity). Within the uncertainties, all the selected complexes but two are consistent with the line fits or above the envelope (Fig.~\ref{fig:sigma_halfa}). Interestingly, these two outlying complexes are consistent with the radius-velocity dispersion correlation. Although the values for another complex are below the envelope, we note that its \ha luminosity was not corrected because its metallicity is unknown. Thus, no complex fails both criteria at the same time. According to these results, self-gravitation should be ensured for more than a half of the complexes for which a determination of the velocity dispersion was achieved. For the remainder, self-gravitation is neither guaranteed nor excluded because at least they generally satisfy one of the two criteria used.

\subsubsection{Dynamical mass estimates}
\label{sub:dyn_mass}

Upper limits to the dynamical masses (\mdyn\twospace) of the selected complexes were derived under the following assumptions: (i) the systems are spherically symmetric; (ii) they are gravitationally bound; and (iii) they have isotropic velocity distributions [$\sigma^2$(total)=3$\times\sigma_{LOS}^2$], where $\sigma_{LOS}$ is the line-of-sight velocity dispersion. In the previous section, we have seen that more than a half of these complexes are likely to be gravitationally bound.

The mass of a virialized stellar system is given by two parameters, its velocity dispersion ($\sigma$) and its half mass radius (r$_{hm}$). The half-mass radius is not an observable and cannot be measured directly. It has to be inferred indirectly by measuring the half-light radius (i.e., \reff\twospace). For a cluster-like object this mass is given by

\begin{equation}
\rm{M}_{\rm{dyn}}/\rm{M}_{\odot}=\eta \frac{\textit{r}_{\rm{eff}}\times \sigma_{LOS}^2}{G}~\rm{,}
\label{eq:dyn_mass1}
\end{equation}
where the \reff is given in pc, $\sigma_{LOS}$ in km s$^{-1}$,  the gravitational constant G=4.3$\times10^{-3}$ pc~\msun\onespace$^{-1}$ $(\rm{km} \rm{s}^{-1})^2$ (\citealt{Spitzer87}), and \mbox{$\eta$ = 9.75} for a wide range of light profiles. However, some studies have shown that $\eta$ is not a constant, and can vary with time (\mbox{$\eta \sim 3-10$}), depending on, for example, the degree of mass segregation and the binary fraction of the cluster (\citealt{Fleck06,Kouwenhoven08}). Since we do not have this information, we use $\eta$ = 9.75.

Since the star-forming knots have a cluster-like light profile (see \citealt{Miralles-Caballero11}) the use of Eq.~\ref{eq:dyn_mass1} to derive the dynamics of the selected complexes seems appropriate. The corresponding masses given in table~\ref{table:derived_prop}, are between a few 10$^7$ and a few 10$^9$~\msun\twospace, with an uncertainty of a factor of 2-3. 

\subsubsection{Total mass: discrepancy between the stellar and the dynamical masses}

It is unclear where to draw the line between a high-mass young super cluster and a low-mass dwarf galaxy in terms of stellar mass. We derive masses only for young population, and those complexes less massive than 10$^6$~\msun  are comparable to the brightest young clusters or complexes hosting recent star formation in other less luminous interacting galaxies (e.g.,~\citealt{Mengel08,Bastian09}). However, does the estimated mass of the young stellar population represent the total mass of the candidates? Only a few studies have reported stellar masses for \tdgs\twospace. The dynamical mass of our complexes is in the range of the typical total baryonic mass of a \tdg (10$^7$-10$^9$~\msun\twospace; e.g.,~\citealt{Duc04,Higdon06,Hancock09}). However, the ratio of the dynamical to the most massive photometric mass in our complexes is suspiciously high (\mbox{\mdyn\twospace/m$_{[M_I]}\sim$ 50-1000)}. Several factors can explain this ratio.

If we assume that the stellar mass represents the total mass of a given complex, this mass might be underestimated, because:

\begin{itemize}
 \item Although the most recent bursts of stars are responsible for the bulk of the ionization of the gas - hence of the emitted flux- the kinematics of the ionizing gas is affected by those of previous stellar generations. In regions of mixed populations, the light is generally dominated by the young stars and the mass by the older population. As outlined before,~\cite{Sheen09} derived a total stellar mass of 3.1$\times$10$^7$~\msun for one of their TDG candidates, a fraction of no more than 2\% being contained in populations younger than 6 Myr. The mass-to-light (M/L) of young (e.g., 5 Myr) and old (e.g., 1 Gyr) populations differs by about a factor of 100. The \ha to the broad-band luminosity colors of some of our candidates are also consistent with those of composite populations. Our complexes might indeed consist of young star-forming knots and evolved population from the complexes themselves and the parent galaxy that has been accreted by the complex. We estimated the total stellar mass of the complex if practically all the \textit{F814W} flux measured inside the whole complex (minus the flux from the knots) belonged to it and came from a 1 Gyr-old population. This total stellar mass approaches significantly to the dynamical mass measured for most of the complexes.\\
 
 \item The total stellar population does not normally represent the total baryonic mass (gas + stellar), and we do not have access to the mass of the gas. Normally, the efficiency of star formation in molecular clouds is very low (typically about 1\%). In (U)LIRGs the efficiency can be higher, but the mass of a molecular cloud undergoing a starburst episode may still be higher than the mass of the starburst by more than one order of magnitude.  The candidates in~\cite{Hancock09} have total stellar masses of \mbox{1-7$\times$10$^6$~\msun\twospace}, the HI mass being \mbox{6$\times$10$^7$~\msun\onespace}. Studies based on radio \hi observations have found that the gas mass of a TDG with a typical diameter of a few kpcs lies within the range between a few 10$^7$ and several 10$^{9}$~\msun (e.g.,~\citealt{Duc00,Bournaud04,Duc07}). \\

\end{itemize}

On the other hand, if we assume that the dynamical mass represents the total mass of a given complex, we might be overestimating it, because:

\begin{itemize}

 \item In each case, \mdyn represents normally an upper limit to the dynamical mass. For instance, given the complex dynamics in interacting systems, high velocity dispersions might not necessarily indicate high masses, but might alternatively represent tidal-flows induced from the merging process and strong winds from LINER-like regions (\citealt{Colina05,Monreal10}). However, at least for the systems that we previously observed with VIMOS, these strong winds may not affect significantly the determination of the velocity dispersions (\citealt{Monreal10}).

  \item Dark matter might contribute significantly to the total TDG mass. However, \tdgs are unlikely to contain a large amount of dark matter (\citealt{Barnes92b}) because their material is drawn from the spiral disk, while the dark matter is thought to surround the galaxy in an extended halo.
 
\end{itemize}

On the basis of all these arguments, it is clear that the total masses of these complexes is somewhere between the photometric and the dynamical masses. Other independent criteria also show that more than a half of the selected complexes are in virial equilibrium, thus the velocity dispersion traces more likely the dynamical mass rather than other processes (see section~\ref{sub:internal_motions}). Given all the caveats and uncertainties previously considered when determining the dynamical and stellar masses, we assume as a mass criterion that the complexes with photometric mass estimates using the broad-band luminosities (single or composite population) that are compatible with 10$^6$~\msun are likely to have a sufficient total mass to become a tidal dwarf galaxy. This assumption should obviously be verified using multi-wavelength photometric and spectroscopic data. Only six of the selected complexes (plus the three candidates in IRAS 16007+3743) fulfill this mass criterion. Had we assumed that a complex with a dynamical mass compatible with \mbox{10$^8$~\msun} or higher has sufficient mass to become a \tdg\twospace, then 15 of the selected complexes would fulfill the mass criterion.

\subsection{Are the selected \ha complexes unaffected by the forces from the parent galaxy?}
\subsubsection{Tidal forces}

If the complex is massive enough to constitute a dwarf galaxy and gravitationally bound, its fate basically depends on the ratio of its mass to the so-called tidal mass (\citealt{Binney87};~\citealt{MendesO01}). The tidal mass condition will tell us whether it is massive enough to survive the tidal forces exerted by the parent galaxy. The tidal mass is defined as
\begin{equation}
 \rm{M}_{\rm{tid}} = 3 \times \rm{M} \left(\frac{\rm{R}}{\rm{D}}\right)^3~\rm{,}
 \label{eq:tidal_mass}
\end{equation}
where M refers to the mass of the parent galaxy, R is the radius of the complex (here the size of the most luminous  knot in a complex, as derived in MC11), and D is the distance to the parent galaxy. This equation is valid when the size (R) of a certain region  is small in comparison with the distance (D) to the parent galaxy, which is the case for all of the complexes (the ratio ``size/projected distance'' is typically below 0.04). If the tidal mass of an object is lower than the total mass, then the object is unaffected by the forces applied by the parent galaxy.

In general, the gravitational potential of interacting galaxies, as in ULIRGs, is a complex function of the mass distribution of the system, which evolves with time. Nevertheless, as a first approximation, we assumed that the distribution is dominated by the masses of the main bodies of the system. Two different approaches were applied in this study: the gravitational potential depends either on:  i) the nearest galaxy (M$_{\rm{tid}}^{\rm{near}}$) or ii) a point mass in the mass center and the total mass of the system (M$_{\rm{tid}}^{\rm{CM}}$). The highest value obtained using the different approaches was taken to compare with the dynamical mass of the complex. 

In each case, we used the measured projected distances $d_{\rm{near}}$ and $d_{\rm{CM}}$, respectively (see table~\ref{table:phot_prop}). Since the projected distance is always smaller than the real one, we actually estimated upper limits to the real tidal masses. To derive the tidal mass, we need to know the mass of the parent galaxy, which in each case is identified with its dynamical mass. Under the same hypothesis as in section~\ref{sub:dyn_mass} of a virialized system, the dynamical mass of a galaxy-like object can be obtained as\\

\begin{equation}
\rm{M}_{\rm{dyn}}/\rm{M}_{\odot}=\rm{m}10^6 \textit{r}_{\rm{hm}}\times \sigma^2~\rm{,} 
\label{eq:dyn_mass2}
\end{equation}
where, the half-mass radius, r$_{hm}$, is given in kpc, $\sigma$ in km s$^{-1}$ and the factor m is a dimensionless function of the assumed mass distribution and ranges from 1.4 for a King stellar mass distribution that adequately represents ellipticals (\citealt{Bender92};~\citealt{Tacconi02}) to 1.75 for a polytropic sphere with a density index covering a range of values (\citealt{Spitzer87}) and 2.09 for a de Vaucoluleurs mass distribution (\citealt{Combes95}). We assumed that m=1.75, as in~\cite{Colina05}. In each case, we used the average value of the velocity dispersion at the peak and nearby surrounding spaxels, covering about a radius of 1\arcsec. Finally, the effective radii (from which we infer r$_{hm}$)  of the parent galaxies were derived using the \textit{H}-band NICMOS and WFC3 images of the galaxies and fitting their two-dimensional light profiles with GALFIT (\citealt{Peng02};~\citealt{Peng10}). In many cases, the NICMOS images do not cover the whole extent of the galaxies, which meant that we had to use the values given in Arribas et al. (in prep.).

In summary, two values (M$_{\rm{tid}}^{\rm{near}}$ and M$_{\rm{tid}}^{\rm{CM}}$) for the tidal mass were obtained using Eq.~\ref{eq:tidal_mass}, and the maximum value was compared to the dynamical mass of the selected complexes derived in section~\ref{sub:dyn_mass}. The ratio M$_{\rm{dyn}}$/M$_{\rm{tid}}$ is in general $>$ 10 (see table~\ref{table:derived_prop}), which ensures that the complexes are unaffected by the tidal forces exerted by the parent galaxy. Only one complex fails this condition and another one is close to failing because of large uncertainties in its dynamical mass estimate. 

\subsubsection{Escape velocity}

If the complex does not have a large enough relative velocity ($v_{\rm{rel}}$) with respect to the galaxy to escape, it might still fall back towards the center of the system because of the gravitational force exerted by the parent galaxy. It is interesting to consider whether a complex does indeed exceed the effective escape velocity ($v_{\rm{esc}}$). This criterion, however, should not have the same weight as the others since it is subject to many uncertainties: (i) the real distance of the complex is always larger than the projected one; (ii) only one component of the velocity is measured and there is no information about the movements in the plane of the sky; and (iii) two possibilities for the relative movements between the region and the system are always possible, since for a given configuration it is impossible to determine whether the complex is closer or further away from the observer than the mass center. We include this criterion in this study for completeness.

For simplicity, we assume that the gravitational potential is created by a point mass representing the total mass of the system, that is located at the mass center. The ratio $|v_{\rm{esc}}|/v_{\rm{rel}}$ (last column in table~\ref{table:derived_prop}) is smaller than 1 for 7 of the selected complexes out of the 22, that is, they could escape. If we take into account projection effects statistically, a complex escapes if \mbox{$|v_{\rm{rel}}| - |v_{\rm{esc}}~\times ~cos(\pi/4)|~>$ 0}. In this case, an additional complex satisfies the criterion.

\subsection{How common is TDG formation in (U)LIRGs?}
\subsubsection{\tdg candidates in (U)LIRGs}

None of the developed criteria can help us determine wheter a given complex will survive the merging process. Projection effects and observational constraints, especially the need for higher angular resolution spectroscopic data (which would allow us to search for velocity gradients), ensure that it is difficult to assess the fate of the selected complexes. In any case, we can investigate which have the higher probabilities of surviving as a \tdg\twospace, based on the fulfillment of a few or most of the criteria viewed in this study. The results are summarized in table~\ref{table:tests}. 

We assigned different weights to the diverse criteria we studied and derived a probability that a certain complex could survive as a \tdg by adding the weights for the criteria that it achieves. The \textit{mass} criterion is considered to be the most important since, even if it is self-gravitating, if it does not have enough mass the complex could be either a bound super cluster or any other entity rather than a low-mass dwarf galaxy. For this reason, it is assigned a weight of 30\%. The criteria with the least importance (each having a weight of 10\%) correspond to the \ha luminosity and the escape velocity ones, since the former comes from empirical considerations and the latter is the least reliable of all. We assigned a weight of 20\% to the M$_{\rm{tid}}$ versus (vs.) M$_{\rm{dyn}}$ criterion. Finally, we considered a somewhat lower weight (15\%) for each of the self-gravitating criterion ($\sigma$ vs. \reff and $\sigma$ vs. \lha), since they prove the same condition. In practice, the complexes that satisfy most of the criteria have the highest probabilities of being \tdg progenitors.

Once the percentage was computed based on which criteria the complexes fulfill, different probabilities that a complex constitutes a \tdgs were defined: \textit{low} (prob $\leq$ 40\%), \textit{medium} (40 $<$ prob $\leq$ 60), \textit{medium-high} (60 $<$ prob $\leq$ 80), and \textit{high} (prob $>$ 80). Only three complexes (counting IRAS 16007+3743 have a high probability. A total of 6 of the 22 selected complexes (9 of 25 if we include those from IRAS 16007+3743) have \textit{medium-high} or \textit{high} probabilities of being \tdg progenitors, and from now on we consider these as our \tdg candidates.

We detect candidates in LIRGs and ULIRGs, although the three candidates with a high probability are only found in ULIRGs. This suggests that \tdg production may be more efficient in systems with higher infrared luminosities. However, this statistic is of very limited robustness because of the low number of candidates found. If we consider only systems for which IFS data are available and located at a distance (i.e.,~\ld $>$ 130 Mpc) where at least a galactocentric radius of 10-15 kpc is covered with the spectroscopic data, we obtain a production rate of about 0.3 candidates per system for the (U)LIRG class. Other \tdgs have been observed at larger distances ($>$30 kpc;~\citealt{Sheen09};~\citealt{Hancock09}) not covered by the FoV of our IFS data. The detection of bright and blue knots (i.e., \mbox{\mi $>$ -12.5} and \mbox{\mbi $<$ 0.5}) in the tidal tails and their tips in the ACS images of some (U)LIRGs (regions not covered with the IFS data; see Fig. 8 in~\citealt{Miralles-Caballero11}), indicative of a young population more massive than 10$^5$~\msun\twospace, may increase the number of candidates for the (U)LIRG class. 

\subsubsection{Dynamical evolution of the TDG candidates}

The number of candidates in systems undergoing the early phases of the interaction process (i.e., phases I-II and II) is 7, significantly larger than the 2 detected in more evolved systems. Normalized to the total number of systems in early and advanced phases, we obtain 0.5 and 0.13 candidates per system, respectively. Therefore, \tdgs are more likely to be formed during the first phases of the interaction in (U)LIRGs.

To analyze the meaning of this trend we need to consider both the life expectancy of these candidates as well as their detectability during their lifetime. Our result does not mean that the total number of candidates in each phase decreases at all. Although large \hi reservoirs have been found in a few TDGs (\citealt{Duc07}) that would be able of sustaining star formation on a Gyr scale, it is unknown whether these are common. Our candidates are indeed \ha\twospace-selected. We might have been unable to identify candidate objects as old as 20 Myr and older, for which their \ha emission is undetectable. There is a score of knots in the initial sample of 32 galaxies at a projected galactocentric distance larger than 2.5 kpc, with colors (\mbox{\mbi $>$ 1.5}) and luminosities (\mbox{\mi $<$ -15}), such that if the population were about 100-1000 Myr old the stellar mass would be higher than 10$^7$-10$^8$~\msun\twospace, close to or similar to the total stellar mass of observed \tdg candidates.

\cite{Bournaud06} ivestigated a set of 96 N-body simulations of colliding galaxies with various mass ratios and encounter geometries, including gas dynamics and star formation. They investigated the dynamical evolution of the TDG candidates found in the various simulations up to \mbox{t = 2 Gyr} after the first pericenter of the relative orbit of the two galaxies. On the basis of the comparison with a higher resolution simulation of a major merger by~\cite{Bournaud08a}, we can roughly assign different dynamical times after the first pericenter for the different phases of interaction defined in~\cite{Miralles-Caballero11}. In this way, the first phases of the interaction (I-II and III) are likely to occurr during the first 400 Myr of the interaction process. If we consider only the observed candidates in (U)LIRGs during these early phases and these systems for which we have available IFS data that cover at least 10-15 kpc of galactocentric radius we derive a production rate of about 0.6 candidates per system. Although the simulations of~\cite{Bournaud06} continue up to \mbox{t = 2 Gyr}, they fit the number of their \tdg candidates versus time with an exponential decay and a lifetime of 2.5 Gyr. They estimate the number of candidates that survive (long-lived candidates that neither fall back nor lose a large fraction of their mass) after 10 Gyr corresponds to 20\% of the \tdg candidates formed during the early phases of the interaction (i.e., \mbox{t $<$ 500 Myr} after the first pericenter). According to this percentage, we estimate a production rate of about 0.1 (20\% of 0.6) long-lived \tdg candidates per system for the (U)LIRG class. 

The average production rate of 0.1 long-lived \tdgs per system can either become lower or higher depending on follow-up studies with IFS capable of providing higher spatial resolution (i.e., adaptive optics assisted systems) and covering a FoV that will even allow us to study galactocentric distances greater than 30 kpc. Modeling and observations claim that the most prominent \tdgs are found along the tidal tails and in particular at their tips (e.g.,~\citealt{Hibbard95,Duc98,Weilbacher00,Higdon06,Duc07,Bournaud08a,Hancock09}). Thus, it would be unsurprising that when covering the whole field of the local (U)LIRG systems with IFS facilities, more TDG candidates will be found. In addition, kinematic data of higher spatial resolution would allow us to determine more reliably whether the candidates are bound. 

\subsubsection{Implications for TDG formation at high-z}

After estimating the production rate of long-lived \tdg candidates per system for the (U)LIRG class, we can now roughly estimate whether satellites of tidal origin are common, and more specifically, the contribution of \tdgs in the early Universe to the dwarf population we see in the local universe. Local (U)LIRGs represent an appropriate class of objects to study the importance of TDGs at high redshift because: (i) they are major contributors to the star formation rate density at \mbox{z $\sim$ 1-2}~\citep{Perez-Gonzalez05}; and (ii) ULIRGs resemble the sub-millimeter galaxies detected at higher redshifts in the sense that they are merging systems with extremely high rates of star formation (\citealt{Chapman03,Frayer03,Engel10}).

\cite{Okazaki00} studied a scenario in which galaxy interactions and/or merger events act as the dominant formation mechanism of dwarf galaxies in any environment and these interactions occur in the context of the hierarchical structure formation in the Universe. They claimed that all dwarfs in the Universe might have a tidal origin. On the basis of this scenario, their statement assumed a production rate of 1 to 2 tidal dwarfs per merger, which has a lifetime of at least 10 Gyr. The production rate we estimate for long-lived candidates in (U)LIRGs is a factor of 10-20 lower, such that their contribution to the overall dwarf population is only 10-5\%. Therefore, we do not find strong evidence that all the local dwarf galaxy population have a tidal origin. This result is consistent with recent estimates implying that the contribution of TDGs to the overall dwarf population is insignificant (\citealt{Bournaud06,Wen11,Kaviraj11}).

The study by~\cite{Okazaki00} covers a wide range of mass ratios for the interacting systems. In our sample, the systems with more than one nucleus span mass ratios from 1:1.7 to 1:3.6, which is the most favorable range for long-lived TDGs, according to the simulations of~\cite{Bournaud06}. Were high-z interactions to have similar mass ratios, the production rate estimated by \cite{Okazaki00} would be higher than the predicted 1-2 tidal dwarf per merger. Thus, our production rate would be a factor of more than 10-20 higher, and the contribution of TDGs to the overall dwarf population would become even lower than 10-5\%. \cite{Rodighiero11} claimed that mergers at \mbox{z $\sim$ 2} have a very low contribution (\mbox{$\sim$ 10\%}) to the cosmic star formation rate density. Were this result to be confirmed, the contribution of \tdgs to the overall dwarf population would still be lower, to a negligible value.

\section{Conclusions}
\label{sec:conclusions}

We have combined high angular resolution \hst images with spectroscopic data from IFS facilities to characterize \mbox{\ha\twospace-emitting} clumps in an initial sample of 27 (U)LIRGs. Our study has extended the search of TDG candidates by~\cite{Monreal07} by considering both LIRG systems and additional and more sensitive images. In particular, we have detected a total of 31 extranuclear star-forming complexes in 11 (U)LIRGs and characterized the main physical and kinematic properties of these complexes. Using structural, physical, and kinematic information, we have also estimated the stellar and dynamical mass content of the complexes and studied their likelihood of resisting the effects of internal as well as external forces based on different dynamical tracers. With all these parameters, we have identified which complexes have the highest probabilities of being long-lived \tdg candidates in (U)LIRGs.  We draw the following conclusions:

\begin{enumerate}[-]

\item Located at an average projected distance of 9.3 kpc from the nucleus, within the range 3.3-13.4 kpc, the structures of the complexes are generally simple, consisting of one or a few compact star-forming regions (knots) in the \hst images, though a few of them reside in a richer cluster environment. The complexes have typical  \textit{B} luminosities (\mbox{\mb $<$ -10.65}), sizes (from few hundreds of pc to about 2 kpc), and \textit{B - I} colors (\mbox{\mbi\twospace $\lesssim$ 1.0}) that are similar to those observed in giant \hii regions and dwarf-like objects. The relatively high metallicities derived, of normally Z$_{\odot}$-Z$_{\odot}$/3 (independent on the luminosity of the complex), reflect the mixing of the metal content in interacting environments, as observed in a few extragalactic \hii regions and \tdg candidates. \\

\item The measured \ha luminosities of the complexes are comparable to those of extremely luminous \ha complexes in nearby systems, giant extragalactic \hii regions, \tdg candidates, and normal dwarfs. Their sizes seem to be more related to the nature of the system. In our case, many complexes have sizes typical of dwarf-like objects.

\item Twenty-two complexes with very simple structures were selected to study their nature as \tdgs\twospace. 
The stellar masses derived for these complexes using the \ha and broad-band luminosities and equivalent widths of a single burst or a composite population (young burst + 1 Gyr) range from a few 10$^4$ to \mbox{10$^8$~\msun\onespace}. Nevertheless, it is typically below \mbox{10$^{6.5}$~\msun\onespace}, which is the lower limit to the total mass. In contrast, the complexes have dynamical masses that are a factor of 50-1000 higher, which provides an upper limit to the total mass.

\item A total of 9 complexes, namely \tdg candidates, have the highest probabilities of becoming dwarf galaxies, at least up to 1-2 Gyr. They are probably massive enough and satisfy most of our criteria for self-gravitation (i.e., the position in the radius-$\sigma$ and the luminosity-$\sigma$ planes) and resistance to the forces from the parent galaxy. We have found evidence that their formation takes place more often in early phases of the interaction.

\item When we consider only systems for which the IFS data cover a significant fraction of the whole system, the production rate of candidates averages about 0.3 per system. This rate is expected to decrease to 0.1 for a long-lived 10 Gyr dwarf, according to recent galaxy merger simulations. Were this to be the case, fewer than 5-10\% of the general dwarf satellite population could be of tidal origin. 
  
\end{enumerate}

\begin{acknowledgements}
We thank the anonymous referee for useful comments that have helped improve the quality of this manuscript. We thank Javier Rodr\'iguez Zaur\'in, Macarena Garc\'ia Mar\'in and Ana Monreal Ibero for providing the spectral maps used in this study. 

This work has been supported by the Spanish Ministry of Education and Science, under grant BES-2007-16198, projects ESP2005-01480, ESP2007-65475-C02-01 and AYA2010-21161-C02-01.
This research has maded use of the NASA/IPAC Extragalactic Database (NED) which is operated by the Jet Propulsion Laboratory, California Institute of Technology, under contract with the National Aeronautics and Space Administration.
We have made use of observations taken with: the NASA/ESA Hubble Space Telescope, obtained from the data archive at the Space Telescope Science Institute (STScI ), which is operated by the Association of Universities for Research in Astronomy, Inc., under NASA contract NAS 5-26555; the William Herschel Telescope operated on the island of La Palma by the ING in the Spanish Observatorio del Roque de los Muchachos of the Instituto de
Astrof\'isica de Canarias; and the Very Large Telescope, operated at the European Southern
Observatory, Paranal (Chile).

\balance
\end{acknowledgements}

\bibliographystyle{aa}

\begin{thebibliography}{127}
\expandafter\ifx\csname natexlab\endcsname\relax\def\natexlab#1{#1}\fi

\bibitem[{{Alonso-Herrero} {et~al.}(2006){Alonso-Herrero}, {Rieke}, {Rieke},
  {Colina}, {P{\'e}rez-Gonz{\'a}lez}, \& {Ryder}}]{Alonso-Herrero06}
{Alonso-Herrero}, A., {Rieke}, G.~H., {Rieke}, M.~J., {et~al.} 2006, \apj, 650,
  835

\bibitem[{{Amor{\'{\i}}n} {et~al.}(2009){Amor{\'{\i}}n}, {Aguerri},
  {Mu{\~n}oz-Tu{\~n}{\'o}n}, \& {Cair{\'o}s}}]{Amorin09}
{Amor{\'{\i}}n}, R., {Aguerri}, J.~A.~L., {Mu{\~n}oz-Tu{\~n}{\'o}n}, C., \&
  {Cair{\'o}s}, L.~M. 2009, \aap, 501, 75

\bibitem[{{Arribas} {et~al.}(1998){Arribas}, {Carter}, {Cavaller}, {del Burgo},
  {Edwards}, {Fuentes}, {Garcia}, {Herreros}, {Jones}, {Mediavilla}, {Pi},
  {Pollacco}, {Rasilla}, {Rees}, \& {Sosa}}]{Arribas98}
{Arribas}, S., {Carter}, D., {Cavaller}, L., {et~al.} 1998, in Society of
  Photo-Optical Instrumentation Engineers (SPIE) Conference Series, Vol. 3355,
  Society of Photo-Optical Instrumentation Engineers (SPIE) Conference Series,
  ed. S.~{D'Odorico}, 821--827

\bibitem[{{Arribas} {et~al.}(2008){Arribas}, {Colina}, {Monreal-Ibero},
  {Alfonso}, {Garc{\'{\i}}a-Mar{\'{\i}}n}, \& {Alonso-Herrero}}]{Arribas08}
{Arribas}, S., {Colina}, L., {Monreal-Ibero}, A., {et~al.} 2008, \aap, 479, 687

\bibitem[{{Barnes} \& {Hernquist}(1992)}]{Barnes92b}
{Barnes}, J.~E. \& {Hernquist}, L. 1992, \nat, 360, 715

\bibitem[{{Bastian} {et~al.}(2006){Bastian}, {Emsellem}, {Kissler-Patig}, \&
  {Maraston}}]{Bastian06}
{Bastian}, N., {Emsellem}, E., {Kissler-Patig}, M., \& {Maraston}, C. 2006,
  \aap, 445, 471

\bibitem[{{Bastian} {et~al.}(2009){Bastian}, {Trancho}, {Konstantopoulos}, \&
  {Miller}}]{Bastian09}
{Bastian}, N., {Trancho}, G., {Konstantopoulos}, I.~S., \& {Miller}, B.~W.
  2009, \apj, 701, 607

\bibitem[{{Bender} {et~al.}(1992){Bender}, {Burstein}, \& {Faber}}]{Bender92}
{Bender}, R., {Burstein}, D., \& {Faber}, S.~M. 1992, \apj, 399, 462

\bibitem[{{Bingham} {et~al.}(1994){Bingham}, {Gellatly}, {Jenkins}, \&
  {Worswick}}]{Bingham94}
{Bingham}, R.~G., {Gellatly}, D.~W., {Jenkins}, C.~R., \& {Worswick}, S.~P.
  1994, in Society of Photo-Optical Instrumentation Engineers (SPIE) Conference
  Series, Vol. 2198, Society of Photo-Optical Instrumentation Engineers (SPIE)
  Conference Series, ed. {D.~L.~Crawford \& E.~R.~Craine}, 56--64

\bibitem[{{Binney} \& {Tremaine}(1987)}]{Binney87}
{Binney}, J. \& {Tremaine}, S. 1987, {Galactic dynamics}, ed. {Binney, J.~\&
  Tremaine, S.}

\bibitem[{{Bournaud} {et~al.}(2004){Bournaud}, {Duc}, {Amram}, {Combes}, \&
  {Gach}}]{Bournaud04}
{Bournaud}, F., {Duc}, P., {Amram}, P., {Combes}, F., \& {Gach}, J. 2004, \aap,
  425, 813

\bibitem[{{Bournaud} {et~al.}(2008){Bournaud}, {Duc}, \&
  {Emsellem}}]{Bournaud08a}
{Bournaud}, F., {Duc}, P., \& {Emsellem}, E. 2008, \mnras, 389, L8

\bibitem[{{Bournaud} \& {Duc}(2006)}]{Bournaud06}
{Bournaud}, F. \& {Duc}, P.-A. 2006, \aap, 456, 481

\bibitem[{{Bresolin} \& {Kennicutt}(1997)}]{Bresolin97}
{Bresolin}, F. \& {Kennicutt}, Jr., R.~C. 1997, \aj, 113, 975

\bibitem[{{Bushouse} {et~al.}(2002){Bushouse}, {Borne}, {Colina}, {Lucas},
  {Rowan-Robinson}, {Baker}, {Clements}, {Lawrence}, \& {Oliver}}]{Bushouse02}
{Bushouse}, H.~A., {Borne}, K.~D., {Colina}, L., {et~al.} 2002, \apjs, 138, 1

\bibitem[{{Cair{\'o}s} {et~al.}(2003){Cair{\'o}s}, {Caon}, {Papaderos},
  {Noeske}, {V{\'{\i}}lchez}, {Garc{\'{\i}}a Lorenzo}, \&
  {Mu{\~n}oz-Tu{\~n}{\'o}n}}]{Cairos03}
{Cair{\'o}s}, L.~M., {Caon}, N., {Papaderos}, P., {et~al.} 2003, \apj, 593, 312

\bibitem[{{Cair{\'o}s} {et~al.}(2001){Cair{\'o}s}, {Caon}, {V{\'{\i}}lchez},
  {Gonz{\'a}lez-P{\'e}rez}, \& {Mu{\~n}oz-Tu{\~n}{\'o}n}}]{Cairos01}
{Cair{\'o}s}, L.~M., {Caon}, N., {V{\'{\i}}lchez}, J.~M.,
  {Gonz{\'a}lez-P{\'e}rez}, J.~N., \& {Mu{\~n}oz-Tu{\~n}{\'o}n}, C. 2001,
  \apjs, 136, 393

\bibitem[{{Chapman} {et~al.}(2003){Chapman}, {Windhorst}, {Odewahn}, {Yan}, \&
  {Conselice}}]{Chapman03}
{Chapman}, S.~C., {Windhorst}, R., {Odewahn}, S., {Yan}, H., \& {Conselice}, C.
  2003, \apj, 599, 92

\bibitem[{{Colina} {et~al.}(2005){Colina}, {Arribas}, \&
  {Monreal-Ibero}}]{Colina05}
{Colina}, L., {Arribas}, S., \& {Monreal-Ibero}, A. 2005, \apj, 621, 725

\bibitem[{{Combes} {et~al.}(1995){Combes}, {Boisse}, {Mazure}, {Blanchard}, \&
  {Seymour}}]{Combes95}
{Combes}, F., {Boisse}, P., {Mazure}, A., {Blanchard}, A., \& {Seymour}, M.
  1995, {Galaxies and Cosmology}, ed. {Combes, F., Boisse, P., Mazure, A.,
  Blanchard, A., \& Seymour, M. }

\bibitem[{{Cresci} {et~al.}(2010){Cresci}, {Mannucci}, {Maiolino}, {Marconi},
  {Gnerucci}, \& {Magrini}}]{Cresci10}
{Cresci}, G., {Mannucci}, F., {Maiolino}, R., {et~al.} 2010, \nat, 467, 811

\bibitem[{{Cui} {et~al.}(2001){Cui}, {Xia}, {Deng}, {Mao}, \& {Zou}}]{Cui01}
{Cui}, J., {Xia}, X.-Y., {Deng}, Z.-G., {Mao}, S., \& {Zou}, Z.-L. 2001, \aj,
  122, 63

\bibitem[{{Dabringhausen} {et~al.}(2008){Dabringhausen}, {Hilker}, \&
  {Kroupa}}]{Dabringhausen08}
{Dabringhausen}, J., {Hilker}, M., \& {Kroupa}, P. 2008, \mnras, 386, 864

\bibitem[{{Denicol{\'o}} {et~al.}(2002){Denicol{\'o}}, {Terlevich}, \&
  {Terlevich}}]{Denicolo02}
{Denicol{\'o}}, G., {Terlevich}, R., \& {Terlevich}, E. 2002, \mnras, 330, 69

\bibitem[{{D{\'{\i}}az} {et~al.}(2007){D{\'{\i}}az}, {Terlevich},
  {Castellanos}, \& {H{\"a}gele}}]{Diaz07}
{D{\'{\i}}az}, {\'A}.~I., {Terlevich}, E., {Castellanos}, M., \& {H{\"a}gele},
  G.~F. 2007, \mnras, 382, 251

\bibitem[{{Dubath} \& {Grillmair}(1997)}]{Dubath97}
{Dubath}, P. \& {Grillmair}, C.~J. 1997, \aap, 321, 379

\bibitem[{{Duc} {et~al.}(2004){Duc}, {Bournaud}, \& {Masset}}]{Duc04}
{Duc}, P., {Bournaud}, F., \& {Masset}, F. 2004, \aap, 427, 803

\bibitem[{{Duc} {et~al.}(2007){Duc}, {Braine}, {Lisenfeld}, {Brinks}, \&
  {Boquien}}]{Duc07}
{Duc}, P., {Braine}, J., {Lisenfeld}, U., {Brinks}, E., \& {Boquien}, M. 2007,
  \aap, 475, 187

\bibitem[{{Duc} {et~al.}(2000){Duc}, {Brinks}, {Springel}, {Pichardo},
  {Weilbacher}, \& {Mirabel}}]{Duc00}
{Duc}, P., {Brinks}, E., {Springel}, V., {et~al.} 2000, \aj, 120, 1238

\bibitem[{{Duc} {et~al.}(1997){Duc}, {Brinks}, {Wink}, \& {Mirabel}}]{Duc97}
{Duc}, P.-A., {Brinks}, E., {Wink}, J.~E., \& {Mirabel}, I.~F. 1997, \aap, 326,
  537

\bibitem[{{Duc} \& {Mirabel}(1994)}]{Duc94}
{Duc}, P.-A. \& {Mirabel}, I.~F. 1994, \aap, 289, 83

\bibitem[{{Duc} \& {Mirabel}(1998)}]{Duc98}
{Duc}, P.-A. \& {Mirabel}, I.~F. 1998, \aap, 333, 813

\bibitem[{{Edmunds} \& {Pagel}(1984)}]{Edmunds84}
{Edmunds}, M.~G. \& {Pagel}, B.~E.~J. 1984, \mnras, 211, 507

\bibitem[{{Ellison} {et~al.}(2008){Ellison}, {Patton}, {Simard}, \&
  {McConnachie}}]{Ellison08}
{Ellison}, S.~L., {Patton}, D.~R., {Simard}, L., \& {McConnachie}, A.~W. 2008,
  \aj, 135, 1877

\bibitem[{{Elmegreen} {et~al.}(1993){Elmegreen}, {Kaufman}, \&
  {Thomasson}}]{Elmegreen93}
{Elmegreen}, B.~G., {Kaufman}, M., \& {Thomasson}, M. 1993, \apj, 412, 90

\bibitem[{{Engel} {et~al.}(2010){Engel}, {Tacconi}, {Davies}, {Neri}, {Smail},
  {Chapman}, {Genzel}, {Cox}, {Greve}, {Ivison}, {Blain}, {Bertoldi}, \&
  {Omont}}]{Engel10}
{Engel}, H., {Tacconi}, L.~J., {Davies}, R.~I., {et~al.} 2010, \apj, 724, 233

\bibitem[{{Ercolano} {et~al.}(2007){Ercolano}, {Bastian}, \&
  {Stasi{\'n}ska}}]{Ercolano07}
{Ercolano}, B., {Bastian}, N., \& {Stasi{\'n}ska}, G. 2007, \mnras, 379, 945

\bibitem[{{Evans} {et~al.}(2002){Evans}, {Mazzarella}, {Surace}, \&
  {Sanders}}]{Evans02}
{Evans}, A.~S., {Mazzarella}, J.~M., {Surace}, J.~A., \& {Sanders}, D.~B. 2002,
  \apj, 580, 749

\bibitem[{{Farrah} {et~al.}(2003){Farrah}, {Afonso}, {Efstathiou},
  {Rowan-Robinson}, {Fox}, \& {Clements}}]{Farrah03}
{Farrah}, D., {Afonso}, J., {Efstathiou}, A., {et~al.} 2003, \mnras, 343, 585

\bibitem[{{Farrah} {et~al.}(2001){Farrah}, {Rowan-Robinson}, {Oliver},
  {Serjeant}, {Borne}, {Lawrence}, {Lucas}, {Bushouse}, \& {Colina}}]{Farrah01}
{Farrah}, D., {Rowan-Robinson}, M., {Oliver}, S., {et~al.} 2001, \mnras, 326,
  1333

\bibitem[{{Fernandes} {et~al.}(2004){Fernandes}, {de Carvalho}, {Contini}, \&
  {Gal}}]{Fernandes04}
{Fernandes}, I.~F., {de Carvalho}, R., {Contini}, T., \& {Gal}, R.~R. 2004,
  \mnras, 355, 728

\bibitem[{{Ferreiro} \& {Pastoriza}(2004)}]{Ferreiro04}
{Ferreiro}, D.~L. \& {Pastoriza}, M.~G. 2004, \aap, 428, 837

\bibitem[{{Ferreiro} {et~al.}(2008){Ferreiro}, {Pastoriza}, \&
  {Rickes}}]{Ferreiro08}
{Ferreiro}, D.~L., {Pastoriza}, M.~G., \& {Rickes}, M. 2008, \aap, 481, 645

\bibitem[{{Fleck} {et~al.}(2006){Fleck}, {Boily}, {Lan{\c c}on}, \&
  {Deiters}}]{Fleck06}
{Fleck}, J.-J., {Boily}, C.~M., {Lan{\c c}on}, A., \& {Deiters}, S. 2006,
  \mnras, 369, 1392

\bibitem[{{Forbes} {et~al.}(2011){Forbes}, {Spitler}, {Graham}, {Foster},
  {Hau}, \& {Benson}}]{Forbes11}
{Forbes}, D., {Spitler}, L., {Graham}, A., {et~al.} 2011, ArXiv e-prints

\bibitem[{{Frayer} {et~al.}(2003){Frayer}, {Armus}, {Scoville}, {Blain},
  {Reddy}, {Ivison}, \& {Smail}}]{Frayer03}
{Frayer}, D.~T., {Armus}, L., {Scoville}, N.~Z., {et~al.} 2003, \aj, 126, 73

\bibitem[{{Frayer} {et~al.}(2004){Frayer}, {Chapman}, {Yan}, {Armus}, {Helou},
  {Fadda}, {Morganti}, {Garrett}, {Appleton}, {Choi}, {Fang}, {Heinrichsen},
  {Im}, {Lacy}, \& {Marleau}}]{Frayer04}
{Frayer}, D.~T., {Chapman}, S.~C., {Yan}, L., {et~al.} 2004, \apjs, 154, 137

\bibitem[{{Garc{\'{\i}}a-Mar{\'{\i}}n}
  {et~al.}(2009{\natexlab{a}}){Garc{\'{\i}}a-Mar{\'{\i}}n}, {Colina}, \&
  {Arribas}}]{Garcia-Marin09b}
{Garc{\'{\i}}a-Mar{\'{\i}}n}, M., {Colina}, L., \& {Arribas}, S.
  2009{\natexlab{a}}, \aap, 505, 1017

\bibitem[{{Garc{\'{\i}}a-Mar{\'{\i}}n}
  {et~al.}(2009{\natexlab{b}}){Garc{\'{\i}}a-Mar{\'{\i}}n}, {Colina},
  {Arribas}, \& {Monreal-Ibero}}]{Garcia-Marin09a}
{Garc{\'{\i}}a-Mar{\'{\i}}n}, M., {Colina}, L., {Arribas}, S., \&
  {Monreal-Ibero}, A. 2009{\natexlab{b}}, \aap, 505, 1319

\bibitem[{{Geha} {et~al.}(2003){Geha}, {Guhathakurta}, \& {van der
  Marel}}]{Geha03}
{Geha}, M., {Guhathakurta}, P., \& {van der Marel}, R.~P. 2003, \aj, 126, 1794

\bibitem[{{Genzel} {et~al.}(1998){Genzel}, {Lutz}, {Sturm}, {Egami}, {Kunze},
  {Moorwood}, {Rigopoulou}, {Spoon}, {Sternberg}, {Tacconi-Garman}, {Tacconi},
  \& {Thatte}}]{Genzel98a}
{Genzel}, R., {Lutz}, D., {Sturm}, E., {et~al.} 1998, \apj, 498, 579

\bibitem[{{Hancock} {et~al.}(2009){Hancock}, {Smith}, {Struck}, {Giroux}, \&
  {Hurlock}}]{Hancock09}
{Hancock}, M., {Smith}, B.~J., {Struck}, C., {Giroux}, M.~L., \& {Hurlock}, S.
  2009, \aj, 137, 4643

\bibitem[{{Hernquist}(1992)}]{Hernquist92a}
{Hernquist}, L. 1992, \nat, 360, 105

\bibitem[{{Hibbard} {et~al.}(1994){Hibbard}, {Guhathakurta}, {van Gorkom}, \&
  {Schweizer}}]{Hibbard94}
{Hibbard}, J.~E., {Guhathakurta}, P., {van Gorkom}, J.~H., \& {Schweizer}, F.
  1994, \aj, 107, 67

\bibitem[{{Hibbard} \& {Mihos}(1995)}]{Hibbard95}
{Hibbard}, J.~E. \& {Mihos}, J.~C. 1995, \aj, 110, 140

\bibitem[{{Higdon} {et~al.}(2006){Higdon}, {Higdon}, \& {Marshall}}]{Higdon06}
{Higdon}, S.~J., {Higdon}, J.~L., \& {Marshall}, J. 2006, \apj, 640, 768

\bibitem[{{Hunsberger} {et~al.}(1996){Hunsberger}, {Charlton}, \&
  {Zaritsky}}]{Hunsberger96}
{Hunsberger}, S.~D., {Charlton}, J.~C., \& {Zaritsky}, D. 1996, \apj, 462, 50

\bibitem[{{Hunter} \& {Gallagher}(1986)}]{Hunter86}
{Hunter}, D.~A. \& {Gallagher}, III, J.~S. 1986, \pasp, 98, 5

\bibitem[{{Iglesias-P{\'a}ramo} \& {V{\'{\i}}lchez}(2001)}]{Iglesias-Paramo01}
{Iglesias-P{\'a}ramo}, J. \& {V{\'{\i}}lchez}, J.~M. 2001, \apj, 550, 204

\bibitem[{{Kaviraj} {et~al.}(2011){Kaviraj}, {Darg}, {Lintott}, {Schawinski},
  \& {Silk}}]{Kaviraj11}
{Kaviraj}, S., {Darg}, D., {Lintott}, C., {Schawinski}, K., \& {Silk}, J. 2011,
  ArXiv e-prints

\bibitem[{{Kewley} {et~al.}(2006){Kewley}, {Groves}, {Kauffmann}, \&
  {Heckman}}]{Kewley06}
{Kewley}, L.~J., {Groves}, B., {Kauffmann}, G., \& {Heckman}, T. 2006, \mnras,
  372, 961

\bibitem[{{Knierman} {et~al.}(2003){Knierman}, {Gallagher}, {Charlton},
  {Hunsberger}, {Whitmore}, {Kundu}, {Hibbard}, \& {Zaritsky}}]{Knierman03}
{Knierman}, K.~A., {Gallagher}, S.~C., {Charlton}, J.~C., {et~al.} 2003, \aj,
  126, 1227

\bibitem[{{Kouwenhoven} \& {de Grijs}(2008)}]{Kouwenhoven08}
{Kouwenhoven}, M.~B.~N. \& {de Grijs}, R. 2008, \aap, 480, 103

\bibitem[{{Kroupa}(2002)}]{Kroupa02}
{Kroupa}, P. 2002, Science, 295, 82

\bibitem[{{Kunth} \& {Schild}(1986)}]{Kunth86}
{Kunth}, D. \& {Schild}, H. 1986, \aap, 169, 71

\bibitem[{{Lamers} \& {Cassinelli}(1999)}]{Lamers99}
{Lamers}, H.~J.~G.~L.~M. \& {Cassinelli}, J.~P. 1999, {Introduction to Stellar
  Winds}, ed. {Lamers, H.~J.~G.~L.~M.~\& Cassinelli, J.~P.}

\bibitem[{{LeFevre} {et~al.}(2003){LeFevre}, {Saisse}, {Mancini}, {Brau-Nogue},
  {Caputi}, {Castinel}, {D'Odorico}, {Garilli}, {Kissler-Patig}, {Lucuix},
  {Mancini}, {Pauget}, {Sciarretta}, {Scodeggio}, {Tresse}, \&
  {Vettolani}}]{leFevre03}
{LeFevre}, O., {Saisse}, M., {Mancini}, D., {et~al.} 2003, in Society of
  Photo-Optical Instrumentation Engineers (SPIE) Conference Series, Vol. 4841,
  Society of Photo-Optical Instrumentation Engineers (SPIE) Conference Series,
  ed. M.~{Iye} \& A.~F.~M. {Moorwood}, 1670--1681

\bibitem[{{Leitherer}(1998)}]{Leitherer98}
{Leitherer}, C., ed. 1998, {Stellar astrophysics for the local group : VIII
  Canary Islands Winter School of Astrophysics, eds., Aparicio, C., Herrero,
  A., S{\'a}nchez, F., p527}

\bibitem[{{Leitherer} {et~al.}(1999){Leitherer}, {Schaerer}, {Goldader},
  {Gonz{\'a}lez Delgado}, {Robert}, {Kune}, {de Mello}, {Devost}, \&
  {Heckman}}]{Leitherer99}
{Leitherer}, C., {Schaerer}, D., {Goldader}, J.~D., {et~al.} 1999, \apjs, 123,
  3

\bibitem[{{L{\'o}pez-S{\'a}nchez} \& {Esteban}(2009)}]{Lopez-Sanchez09}
{L{\'o}pez-S{\'a}nchez}, A.~R. \& {Esteban}, C. 2009, \aap, 508, 615

\bibitem[{{L{\'o}pez-S{\'a}nchez} {et~al.}(2004){L{\'o}pez-S{\'a}nchez},
  {Esteban}, \& {Rodr{\'{\i}}guez}}]{Lopez-Sanchez04}
{L{\'o}pez-S{\'a}nchez}, {\'A}.~R., {Esteban}, C., \& {Rodr{\'{\i}}guez}, M.
  2004, \aap, 428, 425

\bibitem[{{Marlowe} {et~al.}(1997){Marlowe}, {Meurer}, {Heckman}, \&
  {Schommer}}]{Marlowe97}
{Marlowe}, A.~T., {Meurer}, G.~R., {Heckman}, T.~M., \& {Schommer}, R. 1997,
  \apjs, 112, 285

\bibitem[{{Martini} \& {Ho}(2004)}]{Martini04}
{Martini}, P. \& {Ho}, L.~C. 2004, \apj, 610, 233

\bibitem[{{Mateo}(1998)}]{Mateo98}
{Mateo}, M.~L. 1998, \araa, 36, 435

\bibitem[{{Mayya}(1994)}]{Mayya94}
{Mayya}, Y.~D. 1994, \aj, 108, 1276

\bibitem[{{Mendes de Oliveira} {et~al.}(2004){Mendes de Oliveira}, {Cypriano},
  {Sodr{\'e}}, \& {Balkowski}}]{Oliveira04}
{Mendes de Oliveira}, C., {Cypriano}, E.~S., {Sodr{\'e}}, Jr., L., \&
  {Balkowski}, C. 2004, \apjl, 605, L17

\bibitem[{{Mendes de Oliveira} {et~al.}(2001){Mendes de Oliveira}, {Plana},
  {Amram}, {Balkowski}, \& {Bolte}}]{MendesO01}
{Mendes de Oliveira}, C., {Plana}, H., {Amram}, P., {Balkowski}, C., \&
  {Bolte}, M. 2001, \aj, 121, 2524

\bibitem[{{Mengel} {et~al.}(2008){Mengel}, {Lehnert}, {Thatte}, {Vacca},
  {Whitmore}, \& {Chandar}}]{Mengel08}
{Mengel}, S., {Lehnert}, M.~D., {Thatte}, N.~A., {et~al.} 2008, \aap, 489, 1091

\bibitem[{{Michel-Dansac} {et~al.}(2008){Michel-Dansac}, {Lambas}, {Alonso}, \&
  {Tissera}}]{Michel-Dansac08}
{Michel-Dansac}, L., {Lambas}, D.~G., {Alonso}, M.~S., \& {Tissera}, P. 2008,
  \mnras, 386, L82

\bibitem[{{Mihos} \& {Bothun}(1998)}]{Mihos98}
{Mihos}, J.~C. \& {Bothun}, G.~D. 1998, \apj, 500, 619

\bibitem[{{Mirabel} {et~al.}(1991){Mirabel}, {Lutz}, \& {Maza}}]{Mirabel91}
{Mirabel}, I.~F., {Lutz}, D., \& {Maza}, J. 1991, \aap, 243, 367

\bibitem[{{Miralles-Caballero} {et~al.}(2011){Miralles-Caballero}, {Colina},
  {Arribas}, \& {Duc}}]{Miralles-Caballero11}
{Miralles-Caballero}, D., {Colina}, L., {Arribas}, S., \& {Duc}, P.-A. 2011,
  \aj, 142, 79

\bibitem[{{Monreal-Ibero} {et~al.}(2007){Monreal-Ibero}, {Colina}, {Arribas},
  \& {Garc{\'{\i}}a-Mar{\'{\i}}n}}]{Monreal07}
{Monreal-Ibero}, A., {Colina}, L., {Arribas}, S., \&
  {Garc{\'{\i}}a-Mar{\'{\i}}n}, M. 2007, \aap, 472, 421

\bibitem[{{Monreal-Ibero} {et~al.}(2010){Monreal-Ibero}, {V{\'{\i}}lchez},
  {Walsh}, \& {Mu{\~n}oz-Tu{\~n}{\'o}n}}]{Monreal10}
{Monreal-Ibero}, A., {V{\'{\i}}lchez}, J.~M., {Walsh}, J.~R., \&
  {Mu{\~n}oz-Tu{\~n}{\'o}n}, C. 2010, \aap, 517, A27+

\bibitem[{{Montuori} {et~al.}(2010){Montuori}, {Di Matteo}, {Lehnert},
  {Combes}, \& {Semelin}}]{Montuori10}
{Montuori}, M., {Di Matteo}, P., {Lehnert}, M.~D., {Combes}, F., \& {Semelin},
  B. 2010, \aap, 518, A56+

\bibitem[{{Nishiura} {et~al.}(2002){Nishiura}, {Shioya}, {Murayama}, {Sato},
  {Nagao}, {Taniguchi}, \& {Sanders}}]{Nishiura02}
{Nishiura}, S., {Shioya}, Y., {Murayama}, T., {et~al.} 2002, \pasj, 54, 21

\bibitem[{{Okazaki} \& {Taniguchi}(2000)}]{Okazaki00}
{Okazaki}, T. \& {Taniguchi}, Y. 2000, \apj, 543, 149

\bibitem[{{Papaderos} {et~al.}(2006){Papaderos}, {Guseva}, {Izotov}, {Noeske},
  {Thuan}, \& {Fricke}}]{Papaderos06}
{Papaderos}, P., {Guseva}, N.~G., {Izotov}, Y.~I., {et~al.} 2006, \aap, 457, 45

\bibitem[{{Peng} {et~al.}(2002){Peng}, {Ho}, {Impey}, \& {Rix}}]{Peng02}
{Peng}, C.~Y., {Ho}, L.~C., {Impey}, C.~D., \& {Rix}, H.-W. 2002, \aj, 124, 266

\bibitem[{{Peng} {et~al.}(2010){Peng}, {Ho}, {Impey}, \& {Rix}}]{Peng10}
{Peng}, C.~Y., {Ho}, L.~C., {Impey}, C.~D., \& {Rix}, H.-W. 2010, \aj, 139,
  2097

\bibitem[{{P{\'e}rez-Gonz{\'a}lez} {et~al.}(2005){P{\'e}rez-Gonz{\'a}lez},
  {Rieke}, {Egami}, {Alonso-Herrero}, {Dole}, {Papovich}, {Blaylock}, {Jones},
  {Rieke}, {Rigby}, {Barmby}, {Fazio}, {Huang}, \& {Martin}}]{Perez-Gonzalez05}
{P{\'e}rez-Gonz{\'a}lez}, P.~G., {Rieke}, G.~H., {Egami}, E., {et~al.} 2005,
  \apj, 630, 82

\bibitem[{{Pilyugin} {et~al.}(2004){Pilyugin}, {V{\'{\i}}lchez}, \&
  {Contini}}]{Pilyugin04}
{Pilyugin}, L.~S., {V{\'{\i}}lchez}, J.~M., \& {Contini}, T. 2004, \aap, 425,
  849

\bibitem[{{Rela{\~n}o} \& {Beckman}(2005)}]{Relanyo05b}
{Rela{\~n}o}, M. \& {Beckman}, J.~E. 2005, \aap, 430, 911

\bibitem[{{Rela{\~n}o} {et~al.}(2005){Rela{\~n}o}, {Beckman}, {Zurita},
  {Rozas}, \& {Giammanco}}]{Relanyo05}
{Rela{\~n}o}, M., {Beckman}, J.~E., {Zurita}, A., {Rozas}, M., \& {Giammanco},
  C. 2005, \aap, 431, 235

\bibitem[{{Rodighiero} {et~al.}(2011){Rodighiero}, {Daddi}, {Baronchelli},
  {Cimatti}, {Renzini}, {Aussel}, {Popesso}, {Lutz}, {Andreani}, {Berta}, \&
  {Cava}}]{Rodighiero11}
{Rodighiero}, G., {Daddi}, E., {Baronchelli}, I., {et~al.} 2011, ArXiv e-prints

\bibitem[{{Rodr{\'{\i}}guez-Zaur{\'{\i}}n}
  {et~al.}(2011){Rodr{\'{\i}}guez-Zaur{\'{\i}}n}, {Arribas}, {Monreal-Ibero},
  {Colina}, {Alonso-Herrero}, \& {Alfonso-Garz{\'o}n}}]{Rodriguez-Zaurin10}
{Rodr{\'{\i}}guez-Zaur{\'{\i}}n}, J., {Arribas}, S., {Monreal-Ibero}, A.,
  {et~al.} 2011, \aap, 527, A60+

\bibitem[{{Rozas} {et~al.}(2006{\natexlab{a}}){Rozas}, {Richer}, {L{\'o}pez},
  {Rela{\~n}o}, \& {Beckman}}]{Rozas06}
{Rozas}, M., {Richer}, M.~G., {L{\'o}pez}, J.~A., {Rela{\~n}o}, M., \&
  {Beckman}, J.~E. 2006{\natexlab{a}}, \aap, 455, 539

\bibitem[{{Rozas} {et~al.}(2006{\natexlab{b}}){Rozas}, {Richer}, {L{\'o}pez},
  {Rela{\~n}o}, \& {Beckman}}]{Rozas06b}
{Rozas}, M., {Richer}, M.~G., {L{\'o}pez}, J.~A., {Rela{\~n}o}, M., \&
  {Beckman}, J.~E. 2006{\natexlab{b}}, \aap, 455, 549

\bibitem[{{Rupke} {et~al.}(2010){Rupke}, {Kewley}, \& {Barnes}}]{Rupke10}
{Rupke}, D.~S.~N., {Kewley}, L.~J., \& {Barnes}, J.~E. 2010, \apjl, 710, L156

\bibitem[{{Rupke} {et~al.}(2008){Rupke}, {Veilleux}, \& {Baker}}]{Rupke08}
{Rupke}, D.~S.~N., {Veilleux}, S., \& {Baker}, A.~J. 2008, \apj, 674, 172

\bibitem[{{Sanders} \& {Mirabel}(1996)}]{Sanders96}
{Sanders}, D.~B. \& {Mirabel}, I.~F. 1996, \araa, 34, 749

\bibitem[{{Schweizer}(1978)}]{Schweizer78}
{Schweizer}, F. 1978, in IAU Symposium, Vol.~77, Structure and Properties of
  Nearby Galaxies, ed. {E.~M.~Berkhuijsen \& R.~Wielebinski}, 279--284

\bibitem[{{Sheen} {et~al.}(2009){Sheen}, {Jeong}, {Yi}, {Ferreras}, {Lotz},
  {Olsen}, {Dickinson}, {Barnes}, {Park}, {Ree}, {Madore}, {Barlow}, \&
  {Conrow}}]{Sheen09}
{Sheen}, Y., {Jeong}, H., {Yi}, S.~K., {et~al.} 2009, \aj, 138, 1911

\bibitem[{{Sirianni} {et~al.}(2005){Sirianni}, {Jee}, {Ben{\'{\i}}tez},
  {Blakeslee}, {Martel}, {Meurer}, {Clampin}, {De Marchi}, {Ford}, {Gilliland},
  {Hartig}, {Illingworth}, {Mack}, \& {McCann}}]{Sirianni05}
{Sirianni}, M., {Jee}, M.~J., {Ben{\'{\i}}tez}, N., {et~al.} 2005, \pasp, 117,
  1049

\bibitem[{{Spitzer}(1987)}]{Spitzer87}
{Spitzer}, L. 1987, {Dynamical evolution of globular clusters}, ed. {Spitzer,
  L.}

\bibitem[{{Surace} {et~al.}(2000){Surace}, {Sanders}, \& {Evans}}]{Surace00}
{Surace}, J.~A., {Sanders}, D.~B., \& {Evans}, A.~S. 2000, \apj, 529, 170

\bibitem[{{Surace} {et~al.}(1998){Surace}, {Sanders}, {Vacca}, {Veilleux}, \&
  {Mazzarella}}]{Surace98}
{Surace}, J.~A., {Sanders}, D.~B., {Vacca}, W.~D., {Veilleux}, S., \&
  {Mazzarella}, J.~M. 1998, \apj, 492, 116

\bibitem[{{Tacconi} {et~al.}(2002){Tacconi}, {Genzel}, {Lutz}, {Rigopoulou},
  {Baker}, {Iserlohe}, \& {Tecza}}]{Tacconi02}
{Tacconi}, L.~J., {Genzel}, R., {Lutz}, D., {et~al.} 2002, \apj, 580, 73

\bibitem[{{Temporin} {et~al.}(2003){Temporin}, {Weinberger}, {Galaz}, \&
  {Kerber}}]{Temporin03}
{Temporin}, S., {Weinberger}, R., {Galaz}, G., \& {Kerber}, F. 2003, \apj, 587,
  660

\bibitem[{{Terlevich} \& {Melnick}(1981)}]{Terlevich81}
{Terlevich}, R. \& {Melnick}, J. 1981, \mnras, 195, 839

\bibitem[{{Torres-Peimbert} {et~al.}(1989){Torres-Peimbert}, {Peimbert}, \&
  {Fierro}}]{Torres-Peimbert89}
{Torres-Peimbert}, S., {Peimbert}, M., \& {Fierro}, J. 1989, \apj, 345, 186

\bibitem[{{Trager} {et~al.}(1993){Trager}, {Djorgovski}, \& {King}}]{Trager93}
{Trager}, S.~C., {Djorgovski}, S., \& {King}, I.~R. 1993, in Astronomical
  Society of the Pacific Conference Series, Vol.~50, Structure and Dynamics of
  Globular Clusters, ed. {S.~G.~Djorgovski \& G.~Meylan}, 347--+

\bibitem[{{Vacca} {et~al.}(1996){Vacca}, {Garmany}, \& {Shull}}]{Vacca96}
{Vacca}, W.~D., {Garmany}, C.~D., \& {Shull}, J.~M. 1996, \apj, 460, 914

\bibitem[{{van Zee} \& {Haynes}(2006)}]{VanZee06}
{van Zee}, L. \& {Haynes}, M.~P. 2006, \apj, 636, 214

\bibitem[{{V{\'a}zquez} \& {Leitherer}(2005)}]{Vazquez05}
{V{\'a}zquez}, G.~A. \& {Leitherer}, C. 2005, \apj, 621, 695

\bibitem[{{Veilleux} {et~al.}(2006){Veilleux}, {Kim}, {Peng}, {Ho}, {Tacconi},
  {Dasyra}, {Genzel}, {Lutz}, \& {Sanders}}]{Veilleux06}
{Veilleux}, S., {Kim}, D., {Peng}, C.~Y., {et~al.} 2006, \apj, 643, 707

\bibitem[{{Veilleux} {et~al.}(1999){Veilleux}, {Kim}, \&
  {Sanders}}]{Veilleux99}
{Veilleux}, S., {Kim}, D., \& {Sanders}, D.~B. 1999, \apj, 522, 113

\bibitem[{{Veilleux} {et~al.}(1995){Veilleux}, {Kim}, {Sanders}, {Mazzarella},
  \& {Soifer}}]{Veilleux95}
{Veilleux}, S., {Kim}, D., {Sanders}, D.~B., {Mazzarella}, J.~M., \& {Soifer},
  B.~T. 1995, \apjs, 98, 171

\bibitem[{{V\'ilchez} \& {Esteban}(1996)}]{Vilchez96}
{V\'ilchez}, J.~M. \& {Esteban}, C. 1996, \mnras, 280, 720

\bibitem[{{Weilbacher} {et~al.}(2003){Weilbacher}, {Duc}, \&
  {Fritze-v.~Alvensleben}}]{Weilbacher03}
{Weilbacher}, P.~M., {Duc}, P.-A., \& {Fritze-v.~Alvensleben}, U. 2003, \aap,
  397, 545

\bibitem[{{Weilbacher} {et~al.}(2000){Weilbacher}, {Duc}, {Fritze
  v.~Alvensleben}, {Martin}, \& {Fricke}}]{Weilbacher00}
{Weilbacher}, P.~M., {Duc}, P.-A., {Fritze v.~Alvensleben}, U., {Martin}, P.,
  \& {Fricke}, K.~J. 2000, \aap, 358, 819

\bibitem[{{Weilbacher} \& {Fritze-v.~Alvensleben}(2001)}]{Weilbacher01}
{Weilbacher}, P.~M. \& {Fritze-v.~Alvensleben}, U. 2001, \aap, 373, L9

\bibitem[{{Weilbacher} {et~al.}(2002){Weilbacher}, {Fritze-v.~Alvensleben},
  {Duc}, \& {Fricke}}]{Weilbacher02}
{Weilbacher}, P.~M., {Fritze-v.~Alvensleben}, U., {Duc}, P., \& {Fricke}, K.~J.
  2002, \apjl, 579, L79

\bibitem[{{Wen} {et~al.}(2011){Wen}, {Zheng}, {Zhao}, \& {Gao}}]{Wen11}
{Wen}, Z.-Z., {Zheng}, X.-Z., {Zhao}, Y.-H., \& {Gao}, Y. 2011, ArXiv e-prints

\bibitem[{{Yoshida} {et~al.}(1994){Yoshida}, {Taniguchi}, \&
  {Murayama}}]{Yoshida94}
{Yoshida}, M., {Taniguchi}, Y., \& {Murayama}, T. 1994, \pasj, 46, L195

\bibitem[{{Yuan} {et~al.}(2010){Yuan}, {Kewley}, \& {Sanders}}]{Yuan10}
{Yuan}, T., {Kewley}, L.~J., \& {Sanders}, D.~B. 2010, \apj, 709, 884

\bibitem[{{Zwicky}(1956)}]{Zwicky56}
{Zwicky}, F. 1956, Ergebnisse der exakten Naturwissenschaften, 29, 344

\end{thebibliography}

\begin{table*}
\hspace{0cm}
\begin{minipage}{1.0\textwidth}
\renewcommand{\footnoterule}{}  \begin{normalsize}
\caption{Photometric properties of the identified star-forming complexes: distances, magnitudes, and sizes \label{table:phot_prop}}
\begin{center}
\begin{tabular}{lcccccccccc}
\hline \hline
   \noalign{\smallskip}
   IRAS & Nuclei &Complex & Number & d$_{near}$ & d$_{CM}$ & M$_{F435}$ & M$_{F435W}$ - M$_{F814W}$ & \reff & $r$ & $r_{\rm{H} \alpha}$ \\
 & \& IP & Number  & of Knots &(kpc)  & (kpc) & &  & (pc) & (pc) & (pc) \\
 (1) & (2)  & (3)  & (4) & (5) & (6) & (7) & (8) & (9) & (10) & (11)  \\
 \hline
   \noalign{\smallskip}
04315-0840& 1,IV& 1& 4& 4.6& \ldots & -11.37& 1.02& 38& 166& 386\\
& & 2& 1& 4.2& \ldots & -9.34& 0.85& 21& 93& 260\\
06076-2139& 2,III& 1& 2& 11.5& \ldots & -12.86& 0.34& 77& 306& 1216\\
& & 2& 2& 11.3& \ldots & -12.02& 0.21& 59& 283& 902\\
& & 3& 1& 12.6& \ldots & -10.53& 0.29& 53& 121& 384\\
& & 4& 1& 12.3& \ldots & -10.27& 0.26& 51& 116& 544\\
& & 5& 1& 11.1& \ldots & -10.77& -0.23& 41& 89& 608\\
& & 6& 1& 9.9& \ldots & -11.88& 0.35& 66& 137& 942\\
07027-6011 S& 2,I-II& 1& 1& 4.6& \ldots & -11.20& 0.62& 31& 127& 740\\
& & 2& 1& 9.5& \ldots & -10.73& 0.07& 61& 186& 670\\
08355-4944& 1,IV& 1& 2& 3.3& \ldots & -12.94& 0.26& 52& 149& 705\\
08572+3915 N& 2,III& 1& 1& 6.5& 6.8& -12.96& 0.20& 105& 405& 1716\\
08572+3915 SE& 2,III& 1& 5& 4.4& 6.8& -14.54& 0.44& 176& 636& 1514\\
& & 2& 3& 9.3& 11.2& -14.05& 0.39& 199& 560& 1809\\
& & 3& 2& 9.3& 11.2& -12.10& 0.56& 76& 322& 809\\
& & 4& 1& 11.0& 12.1& -11.83& 0.90& 191& 312& 1144\\
F10038-3338& 1,IV& 1& 8& 10.9& \ldots & -14.33& 0.67& 235& 630& 1016\\
& & 2& 4& 8.3& \ldots & -12.47& 0.98& 58& 414& 551\\
& & 3& 2& 12.4& \ldots & -11.34& 0.76& 43& 269& 603\\
& & 4& 2& 13.4& \ldots & -11.72& 0.50& 88& 252& 603\\
12112+0305& 2,III& 1& 3& 3.4& 3.8& -15.56& 1.26& 200& 887& 1721\\
& & 2& 3& 9.4& 9.6& -15.77& 0.56& 212& 506& 1405\\
& & 3& 2& 10.4& 10.8& -14.82& 0.68& 200& 583& 1217\\
& & 4& 1& 11.1& 11.6& -12.34& 0.56& 82& 288& 1405\\
14348-1447& 2,III& 1& 2& 7.8& 7.8& -14.26& 2.34& 280& 909& 2167\\
15250+3609& 1,IV& 1& 4& 8.8& \ldots & -14.13& 1.09& 165& 627& 1176\\
F18093-5744 N& 3,III& 1& 3& 3.9& 3.1& -11.91& 0.56& 48& 170& 497\\
& & 2& 3& 4.8& 6.8& -12.06& 1.11& 46& 178& 602\\
& & 3& 1& 3.8& 6.0& -10.11& 0.86& 20& 74& 363\\
F18093-5744 C& 3,III& 1& 1& 6.5& 4.3& -9.92& 0.84& 37& 78& 482\\
23128-5919& 2,III& 1& 2& 6.8& 8.3& -12.30& 0.87& 83& 376& 1098\\
\hline																							
\noalign{\smallskip}																							
16007+3743	&	3, III	&	R1	&	1	&	16.9	&	17.1	&	\ldots 	&	\ldots 	&	828	&\ldots 	&		1151	\\
	&		&	R2	&	1	&	8.9	&	9.4	&	\ldots 	&	\ldots 	&	884	&\ldots 	&		1375	\\
	&		&	R3	&	1	&	9	&	5.8	&	\ldots 	&	\ldots 	&	851	&\ldots 	&		1562	\\

\hline
\noalign{\smallskip}

\multicolumn{11}{@{} p{\textwidth} @{}}{\textbf{Notes.} Col (1): object designation in the IRAS Point and Faint Source catalogues. In case of several pointings taken with VIMOS data, the orientation (N: North, S: South, SE: South-East, C: Center) of the centered nucleus in that pointing is given. Col(2): Number of nuclei in the system (1-3) and interaction phase (IP) that the system is undergoing according to the classification scheme defined in ~\cite{Miralles-Caballero11}, where I-II indicates first approach, III pre-merger and IV merger phases, respectively. Col (3): identified complex. Col (4): number of knots per complex. Col (5): projected distance to the nearest galaxy. Col (6): projected distance to the mass center of the system. Col (7): Absolute magnitude, M$_{F435}$, not corrected for internal extinction. Col (8): derived photometric color of the complex. Col (9): \reff of the brightest knot inside the complex. Col (10): equivalent radius, derived using the size of the knots (see text). Col (11): radius of the complex measured on the \ha map.}

\end{tabular}
\end{center}
\end{normalsize}
\end{minipage}
\end{table*}

\begin{table*}
\begin{minipage}{\textwidth}
\renewcommand{\footnoterule}{}  \begin{small}
\caption{Spectral observables, metallicity, and dynamical parameters of the identified star-forming complexes \label{table:spec_prop}}
\begin{center}
\begin{tabular}{l@{}c@{\hspace{0.4cm}}c@{\hspace{0.4cm}}c@{\hspace{0.4cm}}c@{\hspace{0.4cm}}c@{\hspace{0.4cm}}c@{\hspace{0.4cm}}c@{\hspace{0.4cm}}c@{\hspace{0.4cm}}c}
\hline \hline
   \noalign{\smallskip}
IRAS & Complex & $F_{\rm{obs}}$ (H$\alpha$) & $L_{\rm{obs}}$ (H$\alpha$) & EW (H$ \alpha$) & Peak   EW & $\frac{Peak~EW}{EW~(H\alpha)}$ & $\sigma$ (\ha\onespace) & $v_{\rm{rel}}$ &Adopted \\
 & Number  & (10$^{-16}$ erg s$^{-1}$ cm$^{-2}$) & (10$^{39}$ erg s$^{-1}$) & (\AA{}) & (\AA{}) &  & (km s$^{-1}$) & (km s$^{-1}$) & 12+log(O/H) \\
 (1) & (2)  & (3)  & (4) & (5) & (6) & (7) & (8) & (9)  & 10\\
 \hline
   \noalign{\smallskip}
04315-0840& 1& 19.3& 1.4& 41.7& 78& 1.9& 16& -26& 8.72\\
& 2& 2.2& 0.2& 8.6& 12& 1.5& 34& -103& 8.99\\
06076-2139& 1& 15.1& 5.8& 48.8& 173& 3.6& 22& 421& 8.50\\
& 2& 5.8& 2.2& 41.3& 80& 2.0& 26& 442& 8.68\\
& 3& 0.6& 0.2& 26.3& 44& 1.7& 23& 442& 8.75\\
& 4& 1.1& 0.4& 27.0& 41& 1.6& 26& 432& 8.53\\
& 5& 1.6& 0.6& 17.8& 28& 1.6& 18& 430& 8.73\\
& 6& 6.8& 2.6& 11.0& 24& 2.2& \ldots & 426& 8.66\\
07027-6011 S& 1& 10.4& 2.9& 68.9& 303& 4.4& 31& 24& 8.65\\
& 2& 2.2& 0.6& \ldots & \ldots & \ldots & \ldots & 73& \ldots \\
08355-4944& 1& 40.8& 40.5& 83.0& 190& 2.3& 30& -25& 8.64\\
08572+3915 N& 1& 7.0& 6.0& 56.6& 242& 4.3& 46& 196& 8.53\\
08572+3915 SE& 1& 15.8& 13.6& 52.2& 183& 3.5& 74& 148& 8.54\\
& 2& 5.8& 5.0& 4.1& 85& 21.0& 62& 207& 8.50\\
& 3& 1.0& 0.8& \ldots & \ldots & \ldots & \ldots & 228& 8.53\\
& 4& 1.0& 0.8& 8.8& 28& 3.2& \ldots & \ldots & 8.56\\
F10038-3338& 1& 12.6& 4.1& 86.0& 374& 4.4& 19& 19& 8.68\\
& 2& 0.6& 0.2& 34.3& 74& 2.2& \ldots & -11& 8.32\\
& 3& 2.0& 0.6& \ldots & \ldots & \ldots & 17& -32& \ldots \\
& 4& 2.2& 0.7& \ldots & \ldots & \ldots & 27& 4& 8.24\\
12112+0305& 1& 44.8& 60.6& 71.1& 252& 3.6& 72& 96& 8.59\\
& 2& 11.9& 16.1& 26.3& 41& 1.6& 81& -204& 8.81\\
& 3& 8.7& 11.8& 11.5& 36& 3.1& \ldots & -349& 8.59\\
& 4& 4.9& 6.7& 6.9& 53& 7.7& \ldots & -266& 8.62\\
14348-1447& 1& 29.5& 65.7& 88.9& 149& 1.7& 54& 106& 8.63\\
15250+3609& 1& 8.1& 5.7& 168.9& 258& 1.5& 69& 124& 8.61\\
F18093-5744 N& 1& 32.8& 2.6& 48.1& 213& 4.4& 23& -76& 8.75\\
& 2& 71.8& 5.6& 67.0& 263& 3.9& 24& 60& 8.77\\
& 3& 9.5& 0.7& 15.8& 21& 1.4& 39& 48& 8.90\\
F18093-5744 C& 1& 3.8& 0.3& 26.0& 57& 2.2& 18& 119& 8.62\\
23128-5919& 1& 5.8& 2.8& 32.0& 51& 1.6& 69& 2& \ldots \\
\hline																			
\noalign{\smallskip}																			
16007+3743	&	R1	&	12	&	57.1	&	40	&	\ldots 	&	\ldots 	&	61	&	-41	&	8.7	\\
	&	R2	&	35.5	&	189.5	&	234	&	\ldots 	&	\ldots 	&	76	&	92	&	8.7	\\
	&	R3	&	5.85	&	8.8	&	80	&	\ldots 	&	\ldots 	&	85	&	338	&	8.7	\\

\hline
\noalign{\smallskip}
\multicolumn{10}{@{} p{\textwidth} @{}}{\textbf{Notes.} Col (1): object designation as in table~\ref{table:phot_prop}. Col (2): identified complex. Col (3): Observed \ha flux. As stated in~\cite{Garcia-Marin09a} and in~\cite{Rodriguez-Zaurin10}, typical uncertainties in the absolute flux calibration are between 15\% and 20\%.  Col (4): derived \ha luminosity with no correction for internal extinction, except for IRAS 16007+3743. Col (5): computed equivalent width (EW) from the \ha and EW maps (see text). Col (6): Spaxel with the largest value of the EW. Col (7): ratio of the peak of the EW to the derived EW for the whole complex. Col (8): central velocity dispersion. Col (9): relative velocity with respect to the  mass center of the system. Col (10): adopted metallicity for the complex, derived using two different line-ratio calibrators (see text).}
\end{tabular}
\end{center}
\end{small}
\end{minipage}
\end{table*}
 
\begin{sidewaystable*}
\vspace{15cm}
\begin{minipage}{\textwidth}
\renewcommand{\footnoterule}{}  \begin{small}
\caption{Derived characteristics of the stellar populations and dynamics of the identified star-forming complexes \label{table:derived_prop}}
\begin{center}
\begin{tabular}{l@{}c@{\hspace{0.3cm}}c@{\hspace{0.3cm}}c@{\hspace{0.3cm}}c@{\hspace{0.3cm}}c@{\hspace{0.3cm}}c@{\hspace{0.3cm}}c@{\hspace{0.3cm}}c@{\hspace{0.3cm}}c@{\hspace{0.3cm}}c@{\hspace{0.3cm}}c@{\hspace{0.3cm}}c@{\hspace{0.3cm}}c@{\hspace{0.3cm}}c@{\hspace{0.3cm}}c@{\hspace{0.3cm}}c}
\hline \hline
   \noalign{\smallskip}
IRAS & Complex & Age$_{\rm{phot}}$ & Age$_{\rm{EW}}$ &  \av & M$_{[I]}$ & M$_{[H_{\alpha}]}$ & Age$_{\rm{Cyoung}}$ & \av\twospace$_C$ & M$_{\rm{Cold}}$ & M$_{\rm{Cyoung}}$  & M$_{\rm{dyn}}$ & M$_{\rm{tid}}^{\rm{near}}$ & M$_{\rm{tid}}^{\rm{CM}}$ & M$_{\rm{dyn}}/\rm{M}_{\rm{tidal}}$ & $v_{\rm{esc}}$ & $|v_{\rm{esc}}|/v_{\rm{rel}}$ \\
\noalign{\smallskip}
 & Number  & (Myr)  & (Myr) & (mags) &  & & (Myr) & (mags) & &  & &  & & (km s$^{-1})$ &  \\
 (1) & (2)  & (3)  & (4) & (5) & (6) & (7) & (8) & (9) & (10) & (11) & (12) & (13) & (14) & (15) & (16) & (17) \\

 \hline
   \noalign{\smallskip}
04315-0840& 1& 3.4& 6.0& 1.7 $\pm$ 0.3& 4.9 $\pm$ 0.1& 5.4 $\pm$ 0.3& 2.1 & 0.2 $\pm$ 0.4& 5.8 $\pm$ 0.1& 3.8 $\pm$ 0.1& 7.5 $\pm$ 0.3 & 6.6 $\pm$ 0.3 & \ldots & 8.1 & 224& 8.6\\
& 2& 4.0& 10.0& 1.5 $\pm$ 0.4& 4.2 $\pm$ 0.2& 5.8 $\pm$ 0.5& 2.8 & 0.7 $\pm$ 0.5& 5.5 $\pm$ 0.2& 3.5 $\pm$ 0.2& 7.8 $\pm$ 0.3 & 6.0 $\pm$ 0.3 & \ldots & $>$ 10 & 235& 2.3\\
06076-2139& 1& 2.9& 5.4& 1.1 $\pm$ 0.3& 5.4 $\pm$ 0.2& 5.9 $\pm$ 0.2& \ldots & \ldots & \ldots & \ldots & 8.0 $\pm$ 0.5 & 5.8 $\pm$ 0.2 & \ldots & $>$ 10 & 96& 0.2\\
& 2& 3.0& 6.0& 1.0 $\pm$ 0.4& 4.9 $\pm$ 0.2& 5.4 $\pm$ 0.3& \ldots & \ldots & \ldots & \ldots & 8.0 $\pm$ 0.3 & 5.7 $\pm$ 0.2 & \ldots & $>$ 10 & 96& 0.2\\
& 3& 4.4& 6.4& 0.6 $\pm$ 0.3& 4.6 $\pm$ 0.1& 5.1 $\pm$ 0.2& 2.9 & 0.0 $\pm$ 0.0& 5.6 $\pm$ 0.1& 3.6 $\pm$ 0.1& 7.9 $\pm$ 0.3 & 4.5 $\pm$ 0.2 & \ldots & $>$ 10 & 92& 0.2\\
& 4& 3.0& 6.5& 0.8 $\pm$ 0.4& 4.5 $\pm$ 0.2& 5.0 $\pm$ 0.5& \ldots & \ldots & \ldots & \ldots & 8.0 $\pm$ 0.3 & 4.5 $\pm$ 0.2 & \ldots & $>$ 10 & 93& 0.2\\
& 5& 2.9& 6.8& 0.4 $\pm$ 0.2& 4.4 $\pm$ 0.2& 5.1 $\pm$ 0.4& \ldots & \ldots & \ldots & \ldots & 7.6 $\pm$ 0.4 & 4.3 $\pm$ 0.2 & \ldots & $>$ 10 & 97& 0.2\\
& 6& 2.8& 7.0& 1.0 $\pm$ 0.3& 5.2 $\pm$ 0.1& 5.9 $\pm$ 0.8& \ldots & \ldots & \ldots & \ldots & \ldots & 5.0 $\pm$ 0.2 & \ldots & \ldots & 103& 0.2\\
07027-6011 S& 1& 2.1& 5.0& 1.2 $\pm$ 0.3& 4.9 $\pm$ 0.2& 5.6 $\pm$ 0.3& \ldots & \ldots & \ldots & \ldots & 7.9 $\pm$ 0.4 & 6.1 $\pm$ 0.3 & \ldots & $>$ 10 & 180& 7.5\\
& 2& 3.2& \ldots & 0.6 $\pm$ 0.3& 4.5 $\pm$ 0.1& 4.6 $\pm$ 0.2& \ldots & \ldots & \ldots & \ldots & \ldots & 5.6 $\pm$ 0.3 & \ldots & \ldots & 125& 1.7\\
08572+3915 N& 1& 2.8& 5.2& 0.7 $\pm$ 0.3& 5.4 $\pm$ 0.2& 5.9 $\pm$ 0.2& \ldots & \ldots & \ldots & \ldots & 8.7 $\pm$ 0.3 & 7.2 $\pm$ 0.4 & 7.4 $\pm$ 0.3 & $>$ 10 & 233& 1.2\\
08572+3915 SE& 3& 4.9& \ldots & 0.9 $\pm$ 0.3& 5.1 $\pm$ 0.1& 5.1 $\pm$ 0.2& 3.1 & 0.0 $\pm$ 0.6& 6.1 $\pm$ 0.2& 4.1 $\pm$ 0.2& \ldots & 6.6 $\pm$ 0.4 & 6.6 $\pm$ 0.3 & \ldots & 225& 0.99\\
& 4& 4.9& 6.9& 1.4 $\pm$ 0.4& 5.5 $\pm$ 0.2& 6.1 $\pm$ 0.2& 3.2 & 0.6 $\pm$ 0.5& 6.5 $\pm$ 0.2& 4.5 $\pm$ 0.2& \ldots & 6.3 $\pm$ 0.4 & 6.5 $\pm$ 0.3 & \ldots & 207& \ldots \\
F10038-3338& 3& 3.6& \ldots & 0.7 $\pm$ 0.3& 4.4 $\pm$ 0.1& 4.5 $\pm$ 0.1& 2.7 & 0.0 $\pm$ 0.0& 5.7 $\pm$ 0.1& 3.7 $\pm$ 0.1& 7.6 $\pm$ 0.3 & 5.8 $\pm$ 0.3 & \ldots & $>$ 10 & 117& 3.6\\
& 4& 4.9& \ldots & 0.9 $\pm$ 0.3& 5.0 $\pm$ 0.1& 5.2 $\pm$ 0.1& 3.0 & 0.0 $\pm$ 0.6& 6.0 $\pm$ 0.2& 4.0 $\pm$ 0.2& 8.2 $\pm$ 0.3 & 5.6 $\pm$ 0.3 & \ldots & $>$ 10 & 113& 26.2\\
12112+0305& 1& 3.4& 5.2& 2.2 $\pm$ 0.3& 7.1 $\pm$ 0.2& 7.3 $\pm$ 0.2& 2.2 & 0.8 $\pm$ 0.4& 8.0 $\pm$ 0.1& 6.0 $\pm$ 0.1& 9.4 $\pm$ 0.2 & 9.2 $\pm$ 0.3 & 9.6 $\pm$ 0.2 & 0.7 & 496& 5.1\\
& 4& 2.1& 6.3& 1.1 $\pm$ 0.3& 5.4 $\pm$ 0.2& 6.1 $\pm$ 0.3& \ldots & \ldots & \ldots & \ldots & \ldots & 6.2 $\pm$ 0.3 & 6.7 $\pm$ 0.2 & \ldots & 275& 1.03\\
14348-1447& 1& 3.3& 5.5& 4.2 $\pm$ 0.3& 7.7 $\pm$ 0.1& 8.1 $\pm$ 0.2& 1.5 & 2.5 $\pm$ 0.3& 8.5 $\pm$ 0.1& 6.5 $\pm$ 0.1& 9.3 $\pm$ 0.2 & 9.1 $\pm$ 1.1 & 9.2 $\pm$ 0.3 & 1.3 & 575& 5.4\\
15250+3609& 1& 5.4& 5.2& 2.4 $\pm$ 0.3& 6.6 $\pm$ 0.1& 6.7 $\pm$ 0.2& 3.9 & 1.1 $\pm$ 0.4& 7.5 $\pm$ 0.1& 5.5 $\pm$ 0.1& 9.3 $\pm$ 0.3 & 7.7 $\pm$ 0.2 & \ldots & $>$ 10 & 214& 1.7\\
F18093-5744 N& 3& 2.9& 7.2& 1.9 $\pm$ 0.3& 5.0 $\pm$ 0.2& 5.8 $\pm$ 0.5& \ldots & \ldots & \ldots & \ldots & 7.9 $\pm$ 0.4 & 4.1 $\pm$ 0.1 & 5.1 $\pm$ 0.2 & $>$ 10 & 220& 4.6\\
F18093-5744 C& 1& 3.5& 6.2& 1.6 $\pm$ 0.3& 4.8 $\pm$ 0.1& 5.2 $\pm$ 0.3& 2.4 & 0.4 $\pm$ 0.4& 5.7 $\pm$ 0.1& 3.7 $\pm$ 0.1& 7.6 $\pm$ 0.5 & 3.6 $\pm$ 0.3 & 5.6 $\pm$ 0.2 & $>$ 10 & 175& 1.5\\
23128-5919& 1& 3.3& 6.3& 1.8 $\pm$ 0.3& 5.7 $\pm$ 0.2& 6.1 $\pm$ 0.5& 2.0 & 0.3 $\pm$ 0.4& 6.5 $\pm$ 0.2& 4.5 $\pm$ 0.2& 9.0 $\pm$ 0.2 & 7.1 $\pm$ 0.2 & 7.0 $\pm$ 0.3 & $>$ 10 & 222& 85.7\\
\hline																																	
\noalign{\smallskip}																																	
16007+3743	&	R1	&	\ldots 	&	7.1	&	1.7	&	8.8	&	\ldots 	&	\ldots 	&	\ldots 	&	\ldots 	&	\ldots 	&	9.8	&	8.2	&	8	&	$>$ 10	&	302	&	7.4	\\
	&	R2	&	\ldots 	&	5.4	&	1.2	&	8.8	&	\ldots 	&	\ldots 	&	\ldots 	&	\ldots 	&	\ldots 	&	10	&	9.2	&	10.1	&	0.9	&	408	&	4.4	\\
	&	R3	&	\ldots 	&	6.4	&	2.3	&	7.8	&	\ldots 	&	\ldots 	&	\ldots 	&	\ldots 	&	\ldots 	&	10.1	&	10	&	9.1	&	1.2	&	519	&	1.5	\\

\hline
\noalign{\smallskip}

\multicolumn{17}{@{} p{\textwidth} @{}}{\textbf{Notes.} Col (1): object designation as in table~\ref {table:phot_prop}. Col (2): identified complex. Col (3): age of the young population, derived using the photometric information (Fig.~\ref{fig:halfa_com}). Col (4): age of the young population using the EW (\ha\twospace). Col (5): internal extinction. Col (6): derived mass of the young population using the I magnitude.  All masses in the table are given in log (M (\msun\twospace)). Col (7): derived mass of the young population  using the \ha flux. Col (8): derived age of the young component, assuming that the candidate consists of a composite population. Col(9): derived internal extinction of the composite population. Col (10): derived mass of the old population (i.e., 1Gyr), assuming a composite population (see text). Col (11): derived mass of the estimated young population, assuming a composite population.  Col (12): dynamical mass. Col (13): tidal mass assuming the potential is created by the nearest galaxy. Col (14): tidal mass assuming the potential is created by a point mass at the mass center of the system. Col (15): ratio of the dynamical to the tidal mass. Col (16): escape velocity. Col (17): ratio of the escape to the relative velocity. }

\end{tabular}
\end{center}
\end{small}
\end{minipage}
\end{sidewaystable*}

\begin{table*}
\hspace{1.2cm}
\begin{minipage}{0.8\textwidth}
\renewcommand{\footnoterule}{}  \caption{Summary of the different criteria used to investigate the nature of the complexes  \label{table:tests}}
\begin{center}
\begin{tabular}{lcccccccc}
\hline \hline
   \noalign{\smallskip}
IRAS	&	Complex	&	Mass	&	L (\ha\twospace)	&	$\sigma$ vs. \reff	&	$\sigma$ vs. \lha	&	M$_{\rm{tid}}$ vs. M$_{\rm{dyn}}$	&	$v_{\rm{esc}}$ vs. $v_{\rm{rel}}$	&	Prob	\\
	&	Number	&		&		&		&		&		&		&		\\
 \hline
   \noalign{\smallskip}
04315-0840	&	1	&	N	&	Y	&	Y	&	Y	&	Y	&	N	&	Medium	\\
	&	2	&	N	&	N	&	N	&	Y	&	Y	&	N	&	Low	\\
06076-2139	&	1	&	N	&	Y	&	Y	&	Y	&	Y	&	Y	&	Medium-high	\\
	&	2	&	N	&	Y	&	Y	&	Y	&	Y	&	Y	&	Medium-high	\\
	&	3	&	N	&	N	&	Y	&	Y	&	Y	&	Y	&	Medium	\\
	&	4	&	N	&	N	&	Y	&	N	&	Y	&	Y	&	Medium	\\
	&	5	&	N	&	Y	&	Y	&	Y	&	Y	&	Y	&	Medium-high	\\
	&	6	&	?	&	Y	&	?	&	?	&	?	&	Y	&	Low	\\
07027-6011 S	&	1	&	N	&	Y	&	N	&	Y	&	Y	&	N	&	Medium	\\
	&	2	&	N	&	Y	&	?	&	?	&	?	&	N	&	Low	\\
08572+3915 N	&	1	&	N	&	Y	&	Y	&	Y	&	Y	&	N	&	Medium	\\
08572+3915 SE	&	3	&	?	&	Y	&	?	&	?	&	Y	&	Y	&	Low	\\
	&	4	&	Y	&	Y	&	?	&	?	&	?	&	?	&	Low	\\
F10038-3338	&	3	&	N	&	Y	&	Y	&	?	&	Y	&	N	&	Medium	\\
	&	4	&	Y	&	Y	&	Y	&	N	&	Y	&	N	&	Medium-high	\\
12112+0305	&	1	&	Y	&	Y	&	N	&	Y	&	N	&	N	&	Medium	\\
	&	4	&	?	&	Y	&	?	&	?	&	?	&	Y	&	Low	\\
14348-1447	&	1	&	Y	&	Y	&	Y	&	Y	&	Y	&	N	&	High	\\
15250+3609	&	1	&	Y	&	Y	&	N	&	Y	&	Y	&	N	&	Medium-high	\\
F18093-5744 N	&	3	&	N	&	Y	&	N	&	Y	&	Y	&	N	&	Medium	\\
F18093-5744 C	&	2	&	N	&	Y	&	Y	&	Y	&	Y	&	N	&	Medium	\\
23128-5919 	&	1	&	Y	&	Y	&	N	&	?	&	N	&	N	&	Low	\\

\hline
\noalign{\smallskip}																	
16007+3743	&	R1	&	Y	&	Y	&	Y	&	Y	&	Y	&	N	&	High	\\
	&	R2	&	Y	&	Y	&	Y	&	Y	&	N	&	N	&	High-medium	\\
	&	R3	&	Y	&	Y	&	Y	&	Y	&	Y	&	N	&	High	\\
\hline
\noalign{\smallskip}
\multicolumn{9}{@{} p{1.1\textwidth} @{}}{\textbf{Notes.} The different criteria are explained throughout the text. The letter Y indicates that the given complex satisfies the criterion, while the letter N indicates that it fails. Symbols with a question mark are either doubtful or indicate that we do not have the data to study the criterion.  The last column indicates the probability that a given complex constitutes a TDG candidate based on these criteria.}
\end{tabular}
\end{center}
\end{minipage}
\end{table*}

\end{document}